\documentclass[12pt]{article}
\usepackage[utf8]{inputenc}
\usepackage{amsmath}
\usepackage{amssymb}
\usepackage{graphicx}
\usepackage{array}

\makeatletter
\newcommand{\oset}[3][0ex]{%
  \mathrel{\mathop{#3}\limits^{
    \vbox to#1{\kern-2\ex@
    \hbox{$#2$}\vss}}}}
\makeatother

\makeatletter

\DeclareTextSymbolDefault{\textquotedbl}{T1}

\numberwithin{equation}{section}

\@ifundefined{date}{}{\date{}}

\usepackage{braket}
\usepackage{bm}
\usepackage{bbm}

\usepackage{epsfig}
\usepackage{pstricks}
\usepackage{color}

\usepackage{hyperref}

\newcommand{\trace}{\text{tr}}
\newcommand{\G}[1]{\text{#1}}
\newcommand{\mycomment}[1]{}
\newcommand{\sectref}[1]{Section \ref{#1}}

\@ifundefined{showcaptionsetup}{}{%
 \PassOptionsToPackage{caption=false}{subfig}}
\usepackage{subfig}
\makeatother

\begin{document}
\begin{titlepage}
\renewcommand{\thefootnote}{\fnsymbol{footnote}}

\begin{flushright} 
KEK-TH-2351
\end{flushright} 


\begin{center}
  {\bf \large Tensor renormalization group and the volume independence\\
    in 2D $\text{U}(N)$ and $\text{SU}(N)$ gauge theories}

\end{center}



\begin{center}
         Mitsuaki H{\sc irasawa}$^{1)}$\footnote
          { E-mail address : mitsuaki@post.kek.jp},
         Akira M{\sc atsumoto}$^{1)}$\footnote
          { E-mail address : akiram@post.kek.jp},\\
         Jun N{\sc ishimura}$^{1,2)}$\footnote
          { E-mail address : jnishi@post.kek.jp} and
         Atis Y{\sc osprakob}$^{1)}$\footnote
          { E-mail address : ayosp@post.kek.jp}


\vspace{1cm}

$^{1)}$\textit{Department of Particle and Nuclear Physics,}\\
\textit{School of High Energy Accelerator Science,}\\
{\it Graduate University for Advanced Studies (SOKENDAI),\\
1-1 Oho, Tsukuba, Ibaraki 305-0801, Japan} 

~

$^{2)}$\textit{KEK Theory Center,
Institute of Particle and Nuclear Studies,}\\
{\it High Energy Accelerator Research Organization,\\
1-1 Oho, Tsukuba, Ibaraki 305-0801, Japan} 
\end{center}

\vspace{0.5cm}

\begin{abstract}
  \noindent The tensor renormalization group method is a promising
  approach to lattice field theories, which is free from the sign problem
  unlike standard Monte Carlo methods.
  One of the remaining issues is the application to gauge theories,
  which is so far limited to U(1) and SU(2) gauge groups.
  In the case of
  higher rank,
  it becomes highly nontrivial to restrict the number
  of representations in the character expansion
  to be used in constructing the fundamental tensor.
  We propose a practical strategy to accomplish this and demonstrate it
  in 2D U($N$) and SU($N$) gauge theories, which are exactly solvable.
  Using this strategy, we obtain
  the singular-value spectrum of the fundamental tensor,
  which turns out to have a definite profile in the large-$N$ limit.
  For the U($N$) case, in particular,
  we show that the large-$N$ behavior of the singular-value spectrum
  changes
  qualitatively at the critical coupling of the Gross-Witten-Wadia
  phase transition.
  As an interesting
  consequence,
  we find a new type of
  volume independence in the large-$N$ limit of the 2D U($N$) gauge theory
with the $\theta$ term in the strong coupling phase,
  which goes beyond the Eguchi-Kawai reduction.
\end{abstract}
\vfill
\end{titlepage}
\vfil\eject


\setcounter{footnote}{0}
\section{Introduction}
Recently the tensor renormalization group method (TRG) has attracted a lot of attention
as a new method to investigate lattice field theories \cite{Levin2007}.
In particular, it is totally free from the
notorious
sign problem
due to its non-stochastic nature unlike conventional Monte Carlo methods.
Another remarkable feature of the method is its accessibility to the infinite volume limit;
the computational cost grows only logarithmically with the lattice size
as opposed to Monte Carlo methods whose cost grows at least linearly.
This owes to the efficient coarse-graining procedure reminiscent of the renormalization group
based on the singular-value decomposition of the fundamental tensor.

The method was originally
applied to the 2D Ising model \cite{Levin2007},
and it was extended to other spin models \cite{Evenbly2015,Yu2014,Xie2012,Wang2014}
and theories with continuous variables such as
scalar field theories \cite{Shimizu2012,Kadoh2019}.
As yet another notable feature of the TRG, it allows
direct implementation
of fermionic degrees of freedom as Grassmann
variables \cite{Shimizu2014,Shimizu2014a,Takeda2015,Sakai:2017jwp,Shimizu:2017onf,Kadoh2018,Yoshimura:2017jpk,Butt:2019uul,Akiyama:2020sfo,Akiyama:2020soe,Akiyama:2021xxr}
unlike in Monte Carlo methods, which inevitably require some sort of ``bosonization''
leading to huge increase in the computational cost.
While extension to higher dimensional theories is not as straightforward as in Monte Carlo methods,
there are various proposals for efficient schemes
to construct a coarse-grained tensor
network \cite{Adachi2019,Kadoh2019a}, which have been
successfully applied to
simple four-dimensional theories \cite{Akiyama:2019xzy,Akiyama:2020ntf,Akiyama:2021zhf}.

One of the remaining issues in the TRG is the application to gauge theories,
which is so far limited to U(1) and $\G{SU}(2)$
gauge groups \cite{Kuramashi:2019cgs,Bazavov2015,UnmuthYockey2018,Bazavov2019}.
In view of this situation,
%
%
here we discuss its application to
$\G{U}(N)$ and $\G{SU}(N)$ gauge theories
using the character expansion
to rewrite the group integral as a sum over discrete indices.
Unlike in U(1) and SU(2) gauge theories,
it is highly nontrivial to restrict the number of representations to be
used in constructing the fundamental tensor.
We propose a practical strategy to accomplish this, and
apply it to the 2D case,
  which is exactly solvable \cite{Gross1980,Wadia1980,Rusakov1990}.
It turns out
that all the procedures for the TRG can be
worked out explicitly.
%
Using the proposed strategy for restricting the number of representations,
we are able to obtain the singular-value spectrum
efficiently.
%
We find, in particular, that the singular-value spectrum
of the fundamental tensor thus obtained
  has a definite profile in the large-$N$ limit.
  Based on this fact, we propose
  a novel interpretation of the volume independence in the large-$N$ limit
  known as
  the Eguchi-Kawai reduction \cite{Eguchi1982}
  in terms of
  the tensor renormalization group.

  We also discuss how the expectation values of the observables
  depend on the cutoff $D_{\rm cut}$
  in the singular-value spectrum.
  In particular, we show that the finite $D_{\rm cut}$ effects
  become severe
  for small $N$, small volume and at weak coupling.
  The proposed strategy enables us to obtain explicit results even in such cases.
  We show how the well-known Gross-Witten-Wadia third-order phase transition
  \cite{Gross1980,Wadia1980} appears as $N$ is increased.
  We also obtain explicit results for the $\G{U}(N)$ gauge theory
  with the $\theta$ term,
  which are inaccessible
  to standard Monte Carlo methods due to the severe sign problem.
  We show how the first-order phase transition at $\theta=\pi$ associated with
  the spontaneous breaking of parity symmetry appears as the volume is increased.
  Moreover, we find a new type of
  volume independence in the large-$N$ limit of the 2D U($N$) gauge theory
  with the $\theta$ term in the strong coupling phase,
  which goes beyond the
  Eguchi-Kawai reduction.
  We provide a theoretical understanding of this property
  by investigating the large-$N$ behavior of the singular-value spectrum
  in the presence of the $\theta$ term.



  
The rest of this paper is organized as follows.
In \sectref{2dUNplusTheta}, we define $\G{U}(N)$ and $\G{SU}(N)$
lattice gauge theories in two dimensions,
and construct the fundamental tensor for the TRG.
In \sectref{TRGapproach}, 
we briefly review the TRG in 2D and explain, in particular,
the coarse-graining procedure.
In \sectref{SVAnalysis}, we propose a practical strategy
to restrict the number of representations
to be used in constructing the fundamental tensor.
We obtain the singular-value spectrum of the fundamental tensor,
and discuss the large-$N$ behavior
and its implications including
the novel interpretation of the Eguchi-Kawai reduction.
In \sectref{sec:explicit_results},
we present explicit results for observables at finite $N$ and finite volume.
We discuss how
the Gross-Witten-Wadia third-order phase transition
appears as $N$ is increased.
In \sectref{section:ThetaTerm},
we investigate 2D $\G{U}(N)$ gauge theory with the $\theta$ term,
and show how the first-order phase transition appears as the volume is increased.
We also discuss the new type of volume independence
in the large-$N$ limit in the strong coupling phase,
which goes beyond the Eguchi-Kawai reduction.
\sectref{SummaryandOutlook} is devoted to a summary and discussions.
In appendix \ref{sec:proof}, we
prove a mathematical statement
about the dimensionality of irreducible representations, which plays a crucial
role in our strategy for restricting the number of representations.

\paragraph{Note added.}
When this paper was about to be completed,
we encountered a preprint \cite{Fukuma:2021cni}
on the arXiv, in which the TRG is applied to
the SU(2) and SU(3) gauge theories in 2D
without using the character expansion.
The fundamental tensor is made finite-dimensional by
replacing the group integral by
a sum over a finite number of randomly chosen SU($N$) matrices
analogously to how the scalar field theories are dealt with in the TRG.
In this method, it is easier to introduce the matter fields
although it is hard to go to larger $N$.


\section{Tensor network representation of 2D gauge theory}
\label{2dUNplusTheta}
In this section, we consider the 2D
$\G{U}(N)$ and $\G{SU}(N)$
gauge theories and rewrite the partition function in the tensor network representation.
In the continuum, the action is given by
\begin{equation}
    S=\frac{1}{4g^2}\int d^2x \, \trace (F_{\mu\nu})^2 \ , 
    \label{eq:continuum_action}
\end{equation}
where
$F_{\mu\nu}$ is the field strength tensor
\begin{equation}
    F_{\mu\nu}=\partial_\mu A_\nu-\partial_\mu A_\mu -i[A_\mu,A_\nu]
    \label{eq:coninuum_fieldstrength}
\end{equation}
defined
with
the gauge field $A_\mu$.

In order to put the theory on a lattice,
we introduce discrete spacetime points $x^\mu =an^\mu$
labeled by the integer vector $n^\mu$ with a lattice spacing $a$.
We define the link variable $U_{n,\mu} \in$ U($N$) or SU($N$),
which corresponds to the gauge field as $U_{n,\mu} \sim \exp(ia A_{\mu}(x))$,
and consider the action
\begin{equation}
  S=-\frac{N}{\lambda}\sum_n\trace(P_n+P_n^\dagger) \ ,
  \label{plaquette-action}
\end{equation}
where we have defined the plaquette
\begin{align}
  P_{n}&=U_{n,1}U_{n+\hat1,2}U^\dagger_{n+\hat2,1}U^\dagger_{n,2} 
  \label{def-plaquette}
\end{align}
and the dimensionless 't Hooft coupling constant $\lambda=2Ng^2a^2$,
which is fixed when we take the $N\rightarrow \infty$ limit
for fixed $a$.
In the continuum limit $a\rightarrow 0$ with fixed $g^2 N$, one
retrieves the action \eqref{eq:continuum_action} in the continuum.
The partition function of the lattice theory is given as
\begin{align}
  Z=\int DU \, e^{-S} \ ,
    \label{eq:defZ}
\end{align}
where $DU$ represents the Haar measure for the link variables.
The 2D lattice gauge theory is exactly solvable as demonstrated
first in the infinite-volume limit \cite{Gross1980,Wadia1980},
and later also for finite volume \cite{Rusakov1990}.
Below we review the evaluation of the partition function
using the character expansion
method \cite{PhysRevD.20.3311,doi:10.1063/1.524368,doi:10.1063/1.524386}.

The first step of the character expansion is to rewrite
the Boltzmann weight $e^{-S}$
in \eqref{eq:defZ} as a product of functions of each plaquette $P_n$,
\begin{align}
  e^{-S} &= \prod_nf(P_n) \ ,
  \label{eq:plaquette_decomposition}\\
    f(P)&=\exp\left\{\frac{N}{\lambda}\trace(P+P^\dagger)\right\} \ .
    \label{eq:U(N)classfunction}
\end{align}
The function $f(P)$ has a special property
\begin{equation}
  f(P) = f(g^{-1}Pg) \ ,  \qquad \forall g\in\G{U}(N)\;\text{or}\;\G{SU}(N) \ ,
\end{equation}
which
allows
the so-called character expansion
\begin{equation}
    f(P) = \sum_r\tilde\beta_r  \, \trace_r(P) \ ,
\end{equation}
where the sum is taken over all the
irreducible representations\footnote{In what follows,
  we refer to irreducible representations
  simply as representations.} of the gauge group and
the symbol $\trace_r$ implies that the trace is taken with respect to the
representation $r$.
The coefficient $\tilde\beta_r$ of the expansion can be evaluated as
\begin{equation}
    \tilde\beta_r=\int dU f(U) \, \trace_r(U^\dagger) \ .
    \label{eq:character_coefficient_integral}
\end{equation}
Substituting \eqref{eq:U(N)classfunction} into
\eqref{eq:character_coefficient_integral}, one obtains \cite{Drouffe1983}
\begin{align}
    \tilde\beta_r&=\left\{
    \begin{array}{ll}
    \det\mathcal{M}_{r, 0}& \mbox{for~}\G{U}(N) \ ,\\
    \displaystyle\sum_{q\in\mathbb{Z}}\det\mathcal{M}_{r,q}& \mbox{for~}\G{SU}(N) \ .
    \end{array}
    \right.
    \label{eq:character_coefficient}
\end{align}
The matrix $\mathcal{M}_{r,q}$ is defined by
\begin{align}
 \Big( \mathcal{M}_{r, q} \Big) _{ij}  &=
  \int_{-\pi}^{+\pi}\frac{d\phi}{2\pi}\cos \{ (l_j+i-j+q)\phi \}
  \exp\left(\frac{2N}{\lambda}\cos\phi\right)
  =  I_{l_j+i-j+q} \left( \frac{2N}{\lambda} \right)
    \label{eq:character_coefficient_determinant}
\end{align}
for $i,j=1,2, \cdots , N$, where $I_n(x)$ is the modified Bessel function of the first kind.
Here we have labeled the representation $r$
by a set of $N$ integers $\{l_1,\cdots,l_N\}$ satisfying
\begin{align}
     l_1\geq l_2\geq \cdots\geq l_{N} 
    \label{UNrepresentations}
\end{align}
with an extra constraint $l_N =0$ in the $\G{SU}(N)$ case, where
the label $\{ l_i \}$ is commonly represented by
the Young tableau with $l_i$ boxes in
the $i$th row.
The dimensionality of a representation
$r=\{l_i\}$ for either gauge group is given by
\begin{equation}
  d_r=\prod_{1\leq i<j\leq N}\left(1+\frac{ l_i- l_j}{j-i}\right) \ .
  \label{eq:representation_dimensionality}
\end{equation}

Note that
any representation $ r^\text{(U)}=\{l_i ' \}$ of $\G{U}(N)$
can be obtained uniquely from a
representation $r^\text{(SU)}=\{l_i\}$ of $\G{SU}(N)$ by
\begin{align}
  l_i '  &=l_i+q 
  \label{eq:shifting_by_q}
\end{align}
with some integer $q$.
According to the definition of
the label $\{ l_i \}$,
the representation matrix
$ D^{r^\text{(U)}}(g)$ of $g \in {\rm U}(N)$
for the representation $r^\text{(U)}$ is given by
\begin{align}
    \label{D-U-SU-rel}
  D^{r^\text{(U)}}(g) &= e^{i p\theta}D^{r^\text{(SU)}}(\tilde g) \ , \\
  p \equiv \sum_{i=1}^N l_i '  &=\sum_{i=1}^{N-1} l_i+Nq \ ,
  \label{D-U-SU-rel2}
\end{align}
where
$ D^{r^\text{(SU)}}(\tilde{g})$ is the representation matrix of
$\tilde{g} \in {\rm SU}(N)$ for the representation $r^\text{(SU)}$
and $g = e^{i \theta} \tilde{g}$.
The right-hand side of \eqref{D-U-SU-rel} is invariant
under $\tilde{g} \mapsto e^{\frac{2\pi i k}{N}}\tilde{g} \in {\rm SU}(N)$ and
$\theta  \mapsto \theta - \frac{2\pi k}{N}$ ($k\in {\bf Z}$) since
\begin{align}
  D^{r^\text{(SU)}}(e^{\frac{2\pi i k}{N}} \tilde{g}) &=
  e^{\frac{2\pi i k}{N} \sum l_i  } D^{r^\text{(SU)}}(\tilde{g}) \ .
    \label{D-SU-prop}
\end{align}
In what follows, we use the notation
\begin{alignat}{2}
  {\rm trv}^{\rm (U)}  &= \{ 0, 0, \cdots , 0, 0 \} \ , & \qquad 
    {\rm trv}^{\rm (SU)}  &= \{ 0, 0, \cdots , 0, 0 \} 
   \ , \nonumber \\
     {\rm fnd}^{\rm (U)}  &= \{ 1, 0, \cdots , 0, 0 \}  \ , & \qquad 
       {\rm fnd}^{\rm (SU)}  &= \{ 1, 0, \cdots , 0,  0 \}  \ ,
  \nonumber \\
  \overline{{\rm fnd}}^{\rm (U)}  &= \{ 0, 0, \cdots , 0, -1 \} \ , & \qquad 
    \overline{{\rm fnd}}^{\rm (SU)}  &= \{ 1, 1, \cdots , 1, 0 \} \ ,
   \nonumber \\
             {\rm adj}^{\rm (U)}  &= \{ 1, 0, \cdots , 0, -1 \} \ ,
             & \qquad 
              {\rm adj}^{\rm (SU)}  &= \{ 2, 1, \cdots , 1, 0 \} \ ,
    \label{example-rep}
\end{alignat}
where ``${\rm trv}$'', ``${\rm fnd}$'' and ``${\rm adj}$''
are abbreviation for
the trivial, fundamental and adjoint representations, respectively,
and the bar like the one in ``$\overline{\rm fnd}$'' above
implies the complex conjugate representation.

Using the character expansion, the partition function becomes
\begin{equation}
    Z=\int DU \prod_nf(P_n)=\sum_{\{r(n)\}}\int DU\prod_n\tilde\beta_{r(n)}\trace_{r(n)}(P_n) \ ,
    \label{eq:expandZ}
\end{equation}
where $r(n)$ is the representation that appears
in the character expansion of $f(P_n)$.
Next we decompose $\trace_{r(n)}(P_n)$ in terms of link variables as
\begin{equation}
    \trace_{r(n)}(P_n)= D^{r(n)}_{\alpha\beta}(U_{n,1})
    D^{r(n)}_{\beta\gamma}(U_{n+\hat1,2})
    D^{r(n)}_{\gamma\delta}(U^\dagger_{n+\hat2,1})
    D^{r(n)}_{\delta\alpha}(U^\dagger_{n,2}) \ ,
    \label{plaquette-decomposition}
\end{equation}
where $D^r(U)$ is the representation matrix of $U$ for the representation $r$,
and the indices $\alpha,\beta,\gamma,\delta$
are summed over implicitly.
It can then be seen\footnote{Note that this is the case only in two dimensions.}
that any given link variable $U_{n,\mu}$ appears only twice
in \eqref{eq:expandZ}; one as $U$ and the other as $U^\dagger$.
Therefore, the partition function factorizes into the integral for each link variable as
\begin{align}
  Z&=\prod_{n,\mu}z^{(n,\mu)} \ ,
  \label{eq:integrateU1}
  \\
    (z^{(n,\mu)})_{rs\alpha\beta\gamma\delta}&=
  \tilde\beta_r^{\frac{1}{4}}\tilde\beta_s^{\frac{1}{4}}
    \int dU_{n,\mu}
    D^r_{\alpha\beta}(U_{n,\mu})D^s_{\gamma\delta}(U^\dagger_{n,\mu})
    =\frac{\tilde\beta_r^\frac{1}{2}}{d_r}\delta_{rs}\delta_{\alpha\delta}\delta_{\beta\gamma} \ ,
    \label{eq:integrateU}
\end{align}
where we have used
the orthogonality relation
\begin{equation}
  \int dUD^r_{\alpha\beta}(U)D^s_{\gamma\delta}(U^\dagger)
  =\frac{1}{d_r}\delta_{rs}\delta_{\alpha\delta}\delta_{\beta\gamma}
    \label{eq:grand_orthogonality_theorem}
\end{equation}
with $d_r$ being the dimensionality of the representation $r=\{l_i\}$
given by \eqref{eq:representation_dimensionality}.

\begin{figure}
\centering
\includegraphics[scale=1.0]{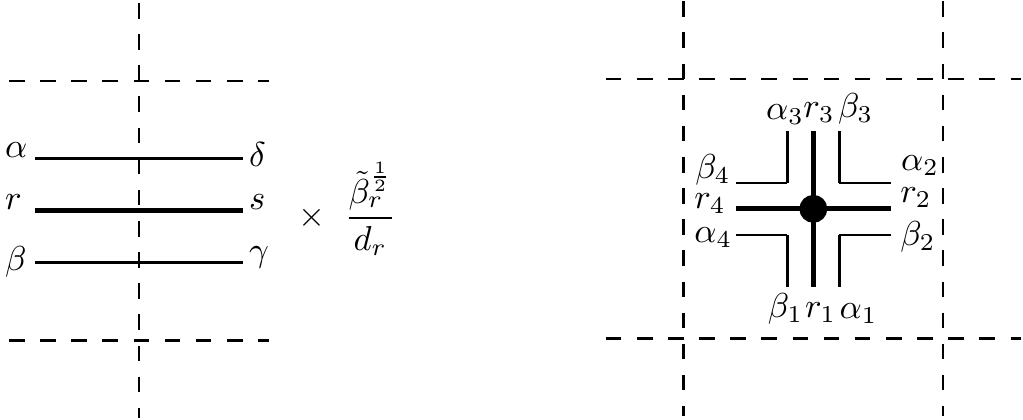}
\caption{The tensor associated with each link (Left)
  and that associated with each plaquette (Right) are depicted
  corresponding to
    \eqref{eq:integrateU} and \eqref{eq:plaquettetensor}, respectively.
}
\label{fig:integrateU}
\end{figure}

\begin{figure}
\centering
\includegraphics[scale=1.0]{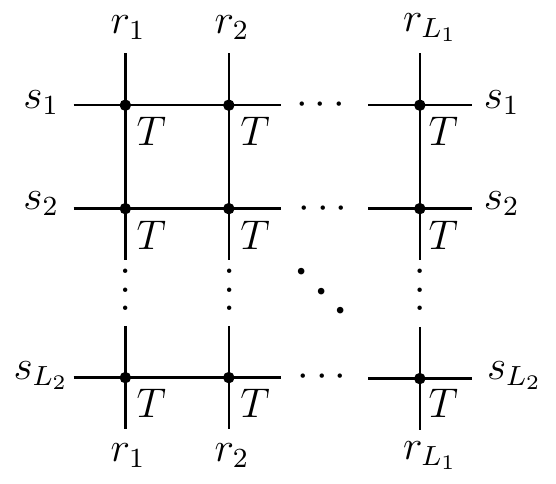}
\caption{The tensor network representing the partition function \eqref{eq:partitionnetwork}.}
\label{fig:tensornetwork}
\end{figure}

The tensor \eqref{eq:integrateU}
can be considered to be associated with each link on the lattice
as depicted in Fig.~\ref{fig:integrateU} (Left).
Note also that on the right-hand side of \eqref{eq:integrateU1},
the indices of
$(z^{(n,\mu)})_{rs\alpha\beta\gamma\delta}$
are assumed to be contracted appropriately 
according to \eqref{plaquette-decomposition}.
As a result, we also have a tensor
\begin{equation}
  \delta_{r_1r_2r_3r_4}\delta_{\alpha_1\beta_2}
  \delta_{\alpha_2\beta_3}\delta_{\alpha_3\beta_4}\delta_{\alpha_4\beta_1} 
    \label{eq:plaquettetensor}
\end{equation}
associated with each plaquette
as depicted in Fig.~\ref{fig:integrateU} (Right),
where we have defined
\begin{align}
\delta_{r_1r_2r_3r_4}   &= \left\{
    \begin{array}{ll}
     1
  & \mbox{for~}  r_1 = \cdots = r_4 \ ,\\
      0
& \mbox{otherwise.}
    \end{array}
    \right.
    \label{def-4way-delta}
\end{align}
Note that the matrix indices form a loop around each site
yielding a factor of $d_r$ for each site.
Thus we end up with
a tensor network
\begin{align}
  Z&=\prod T  \ ,
  \label{eq:partitionnetwork}
\end{align}
where the fundamental tensor $(T)_{pqrs}$ for each plaquette
is defined as
\begin{align}
    (T)_{pqrs}=\frac{\tilde\beta_{p}}{d_p}\delta_{pqrs} \ ,
  \label{eq:fundamental_tensor}
\end{align}
after reassigning the factor $\frac{\tilde\beta_r^\frac{1}{2}}{d_r}$ for each link
to the two plaquettes sharing it and
reassigning the factor $d_r$ for the site $n$ to the plaquette $P_n$.
The indices of the fundamental tensor $(T)_{pqrs}$
in \eqref{eq:partitionnetwork} are assumed to be contracted appropriately
as depicted in Fig.~\ref{fig:tensornetwork}, where we impose
periodic boundary conditions.


In the present case, the structure of the fundamental tensor is
so simple that we can evaluate \eqref{eq:partitionnetwork}
for an $L_1 \times L_2$ lattice as
\begin{equation}
    Z=\sum_r\left(\frac{\tilde\beta_r}{d_r}\right)^{L_1L_2} \ ,
    \label{eq:analytic_partition_function}
\end{equation}
which agrees with the known exact result in Ref.~\cite{Rusakov1990}.
%

\section{Brief review of the TRG in 2D}
\label{TRGapproach}

In this section we briefly review the TRG \cite{Levin2007},
which is a useful method that can be applied to various theories
whose partition function is written
as a network of tensors with translational invariance.
The crucial point is to make use of the coarse graining to
access a large system size efficiently.
Here we explain the method in 2D.

The TRG consists of two steps.
The first step is to decompose the tensor $T$ using the singular-value decomposition (SVD).
Let us consider the following two types of SVD.
The first one amounts to regarding the double indices
$(q,r)$ and $(s,p)$ of the tensor $T_{pqrs}$
as single indices $a$ and $b$, respectively, and applying the SVD to
the resulting matrix with the two indices $a$ and $b$ as
\begin{align}
  T_{pqrs}
  &=
  \sum_{c}
  S^{(1)}_{ac}G^{(1)}_cS^{(2)}_{cb}
  = \sum_{c}
  \tilde{S}^{(1)}_{ac} \tilde{S}^{(2)}_{cb}   \ ,
  \label{eq:SVDplus}
\end{align}
where $S^{(1)}$ and $S^{(2)}$ are unitary matrices and
$G^{(1)}_c$ are the singular values,
which are positive-semidefinite and labeled in the descending order.
In the second equality,
we have absorbed the singular values into the $S^{(1)}$ and $S^{(2)}$ as
\begin{equation}
  \tilde S^{(1)}_{ac}=S^{(1)}_{ac}\sqrt{G^{(1)}_c}  \ ,\quad
  \tilde S^{(2)}_{cb}=S^{(2)}_{cb}\sqrt{G^{(1)}_c} \ .
\end{equation}

Similarly, we define
the second type of SVD by regarding
the double indices $(p,q)$ and $(r,s)$ of the tensor $T_{pqrs}$
as single indices $a$ and $b$, respectively, and applying the SVD to
the resulting matrix with the two indices $a$ and $b$ as
\begin{align}
  T_{pqrs}
  &= \sum_{c}
  S^{(3)}_{ac}G^{(2)}_cS^{(4)}_{cb}
  = \sum_{c}
  \tilde{S}^{(3)}_{ac} \tilde{S}^{(4)}_{cb}  \  ,
  \label{eq:SVDminus}
\end{align}
where $S^{(3)}$, $S^{(4)}$ are unitary matrices and $G^{(2)}_c$ are the singular values,
which are again positive-semidefinite and labeled in the descending order.
In the second equality,
we have absorbed the singular values into
the $S^{(1)}$ and $S^{(2)}$ as
\begin{equation}
  \tilde S^{(3)}_{ac}=S^{(3)}_{ac}\sqrt{G^{(2)}_c} \ ,
  \quad\tilde S^{(4)}_{cb}=S^{(4)}_{cb}\sqrt{G^{(2)}_c} \ .
\end{equation}

\begin{figure}
\centering
\includegraphics[scale=1.0]{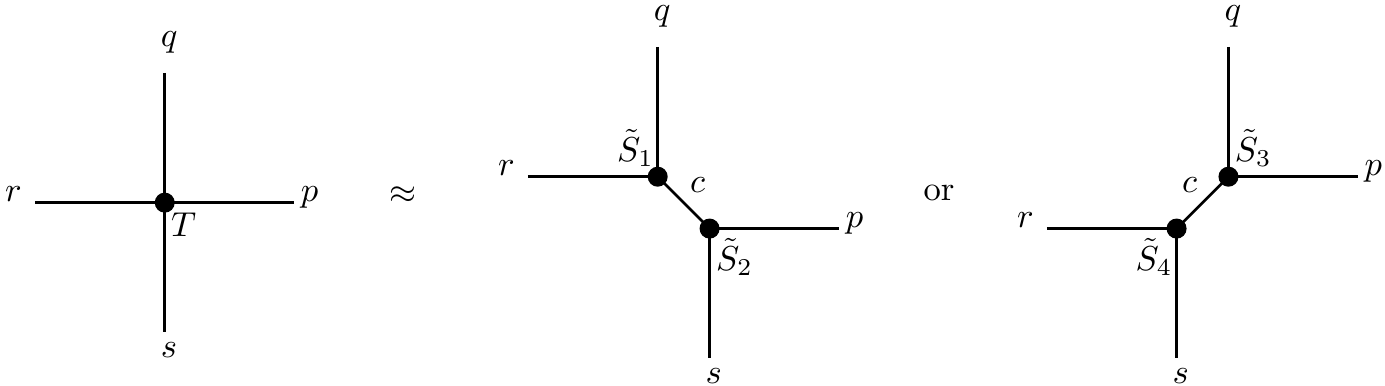}
\caption{The two types of SVD used for even sites and odd sites,
respectively.}
\label{fig:SVD}
\end{figure}

\begin{figure}
\centering
\includegraphics[scale=1.0]{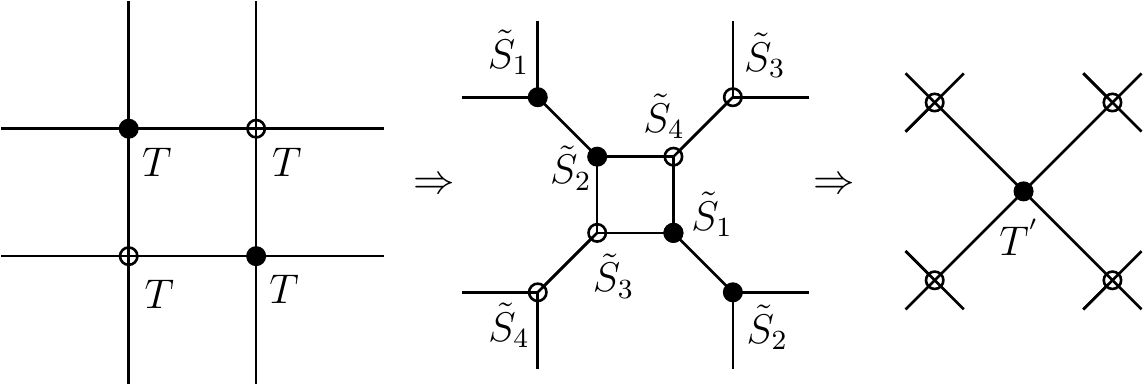}
\caption{The two steps in the coarse-graining procedure for the tensor network.}
\label{fig:decomposed_network}
\end{figure}

The two types of SVD \eqref{eq:SVDplus} and \eqref{eq:SVDminus}
can be represented diagrammatically as in Fig.~\ref{fig:SVD}.
We apply them
on the even and odd sites of the lattice,
respectively\footnote{Here we need to assume that the number of sites
  in each direction
  is even in order to assign the parity to each site consistently to the
  boundary condition.},
which is shown in Fig.~\ref{fig:decomposed_network}
as the first step from the left to the middle.

The second step is to define
\begin{equation}
    T'_{abcd}=\sum_{pqrs}\tilde S^{(1)}_{(q,r),a}\tilde S^{(3)}_{(r,s),d}\tilde S^{(2)}_{c,(s,p)}\tilde S^{(4)}_{b,(p,q)} \ ,
    \label{def-Tprime}
\end{equation}
which represents the coarse-grained version of the original tensor $T_{pqrs}$.
This is shown in Fig.~\ref{fig:decomposed_network}
as the second step from the middle to the right.
Note that the new lattice is tilted by $45^{\circ}$ degrees, but the boundary condition
is such that it respects the translational invariance in the directions of the original lattice.
Repeating this procedure twice, one obtains a $L/2 \times L/2$ lattice
with periodic boundary conditions.
Thus, starting from a $2^n \times 2^n$ lattice,
we can perform the coarse graining $2n$ times to arrive at a one-site model
with the fundamental tensor $T''_{pqrs}$,
whose
partition function can be evaluated as
\begin{equation}
  Z = \sum_{p,q}
  T''_{pqpq} \ .
  \label{part-fn-trg}
\end{equation}

In general, the fundamental tensor has a form more complicated
than \eqref{eq:fundamental_tensor},
and the SVD such as \eqref{eq:SVDplus}, \eqref{eq:SVDminus}
has to be done numerically.
Note also that an index of the coarse-grained version
of the fundamental tensor corresponds to
the double indices of the original one as we discussed above.
Therefore, one needs to truncate the range of the indices
in order to avoid the growth of the numerical cost as the coarse graining proceeds.
The crucial point of the TRG is to
perform this truncation by
keeping only the large singular values
in the expressions like \eqref{eq:SVDplus} and \eqref{eq:SVDminus}.
Thus the fundamental tensor
has indices with the same dimension, which is called the bond dimension $D_{\rm cut}$,
throughout the coarse-graining procedure.
One has to increase
$D_{\rm cut}$ until the results do not change any more.
The required value of $D_{\rm cut}$ is basically determined by
the fall-off of the singular-value spectrum,
which
becomes slow as the correlation length of the theory increases.
The numerical cost of the TRG grows proportionally to $D_{\rm cut}^n$
with some power $n$, which is 6 in the 2D case
and becomes larger for theories in higher dimensions.
Recently, some new TRG schemes \cite{Xie2012,Adachi2019,Kadoh2019a}
for higher dimensional models
have been proposed to reduce
the power $n$ considerably.

\section{Application to U($N$) and SU($N$) gauge theories}
\label{SVAnalysis}

When one applies the TRG to a gauge theory with a continuous group
using the character expansion,
the dimension of the original fundamental tensor
is infinite because there are infinitely many representations.
In the case of U(1) and SU(2) gauge groups, which have been studied so far
in the literature \cite{Kuramashi:2019cgs,Bazavov2015,UnmuthYockey2018,Bazavov2019},
there is a natural choice for restricting the number of representations
because the representation is labeled by the charge and the spin, respectively.
Obviously, this issue becomes more nontrivial in the case of SU($N$) and U($N$) gauge groups with $N\ge 3$.

In the 2D lattice gauge theory that we are considering,
the fundamental tensor
takes a particularly simple form  \eqref{eq:fundamental_tensor},
and one can obtain the product \eqref{eq:partitionnetwork} exactly
to arrive at the well-known result \eqref{eq:analytic_partition_function}.
However, one still has to
truncate the summation over representations
in \eqref{eq:analytic_partition_function}
in order to obtain explicit results
in general.

Here we discuss how we can restrict the number of representations to be included
in the fundamental tensor
efficiently so that the singular-value spectrum is correctly reproduced up to
a given bond dimension $D_{\rm cut}$.
Then, we discuss various properties of the singular-value spectrum,
in particular, at large $N$, and consider their implications.


\subsection{singular-value spectrum in the 2D gauge theories}
\label{sec:sing-val-2D}

In the case of 2D lattice gauge theory,
the fundamental tensor is given by \eqref{eq:fundamental_tensor},
which allows us to make the SVD
trivially.
Namely, for the two types of SVD,
the singular values $G^{(i)}_{(r,s)}$ ($i=1,2$) are given by 
\begin{align}
  G^{(i)}_{(r,s)} = \sigma_r \delta_{rs} \ , \quad \quad
  \sigma_r \equiv \frac{|\tilde\beta _r|}{d_r}  \ .
  \label{def-Gi}
\end{align}
%
One can easily find that the coarse-grained tensor \eqref{def-Tprime} becomes
\begin{align}
    T_{pqrs} ' = (\sigma_p )^2 \, \delta_{pqrs} \ ,
  \label{eq:coarse-grained-tensor}
\end{align}
which has the same form as \eqref{eq:fundamental_tensor}.
Repeating this procedure $2n$ times starting with
a system on the $L \times L$ lattice,
where $L=2^n$,
one obtains a one-site model
with the fundamental tensor
\begin{align}
    T_{pqrs} '' =  (\sigma_p)^{V} \,  \delta_{pqrs} \ ,
  \label{eq:coarse-grained-tensor-final}
\end{align}
where $V=L^2$ is the volume of the lattice. 
Plugging this into \eqref{part-fn-trg},
the partition function
is given in terms of the singular values $\sigma_r$ as
\begin{equation}
  Z = \sum_r (\sigma_r)^V \ ,
    \label{eq:TRGpartition}
\end{equation}
which agrees with the exact result \eqref{eq:analytic_partition_function}.
Thus, the TRG can be worked out explicitly in this model,
and the only remaining task is to determine
the largest $D_\text{cut}$ singular values of the fundamental tensor.

Due to the simple $V$-dependence of the partition function,
one finds
in the infinite-volume limit $V \rightarrow \infty$
that the summation in \eqref{eq:TRGpartition} is dominated
by the largest singular value,
which is known to
correspond to the trivial representation,
\begin{equation}
 \sigma_{\rm trv}=\int dU
  \exp\left\{\frac{N}{\lambda}\trace(U+U^\dagger)\right\} \ ,
    \label{eq:character_coefficient_integral-1}
\end{equation}
where we have used \eqref{eq:character_coefficient_integral} and \eqref{eq:U(N)classfunction}.
The right-hand side of \eqref{eq:character_coefficient_integral-1} is nothing
but the partition function of the one-plaquette model obtained in the $V \rightarrow \infty$ limit
of 2D lattice gauge theory \cite{Gross1980,Wadia1980}, and
in the large-$N$ limit, one obtains
\begin{align}
C^{(0)} \equiv 
\lim_{N \rightarrow \infty} \frac{1}{N^2}  \log \sigma_{\rm trv}
  =\left\{
   \begin{array}{ll}
     \displaystyle
     \frac{2}{\lambda}+\frac{1}{2}\log\frac{\lambda}{2}-\frac{3}{4}&\mbox{~for~}\lambda<2 \ ,\\[2mm]
     \displaystyle \frac{1}{\lambda^2}& \mbox{~for~}\lambda\geq 2
    \end{array}
    \right. 
    \label{eq:GWW_vacuum_energy}
\end{align}
for both SU($N$) and U($N$) gauge theories.
Let us emphasize, however, that this simplification in the $V \rightarrow \infty$ limit
is peculiar to the present 2D gauge theories.
The change of the behavior at $\lambda=2$ in \eqref{eq:GWW_vacuum_energy}
indicates the Gross-Witten-Wadia phase transition,
which plays an important role
in what follows.

\subsection{restricting the number of representations}
\label{chooserep}

Here we discuss how to restrict the number of representations
in the case of SU($N$) and U($N$) gauge groups,
which is nontrivial
since the representations are labeled by more than one parameters $\{l_i\}$.
As a natural requirement, 
the cutoff scheme should not
discriminate representations which are complex conjugate to each other.
We will consider three cutoff schemes which satisfy this requirement
and discuss which is the most efficient.


Let us first consider the $\G{SU}(N)$ case.
As is mentioned in Section \ref{2dUNplusTheta},
the representations of $\G{SU}(N)$ are parametrized by $(N-1)$ non-negative
integers $l_1,l_2, \cdots ,l_{N-1}$
satisfying
\begin{equation}
    l_1\geq l_2\geq \cdots \geq l_{N-1} \geq l_N  = 0 \ .
\end{equation}
Considering that the complex conjugate representation of $\{ l_i \}$ is given
by $\{ \bar{l}_i \}$, where $\bar{l}_i = l_1 - l _{N+1-i}$,
we have $\bar{l}_1 = l_1$.
In view of this, a simple choice would be to introduce 
a cutoff $l_1\leq\Lambda$, where $\Lambda$ is some integer.
The number of representations within this cutoff can be
shown\footnote{Let us introduce
  integers $n_1 , \cdots , n_N \ge 0$
  through $l_{1}= \Lambda - n_1$ and $l_{i+1} = l_i - n_{i+1}$ ($i=1,\cdots, N-1$),
  which implies $l_N = \Lambda - \sum_{j=1}^{N} n_j$.
  Imposing $l_N=0$, we get the relation $\sum_{j=1}^{N}n_j=\Lambda$.
  Thus, the issue of counting the representations
  with $l_1\leq\Lambda$ reduces to that of assigning $\Lambda$ identical objects
  into $N$ partitions.} to be
${}_{\Lambda+N-1} {\rm C} _{\Lambda}$,
which grows as $\Lambda^{N-1}$ with $\Lambda$ for fixed $N$.

\begin{figure}
\centering
\includegraphics[scale=0.85]{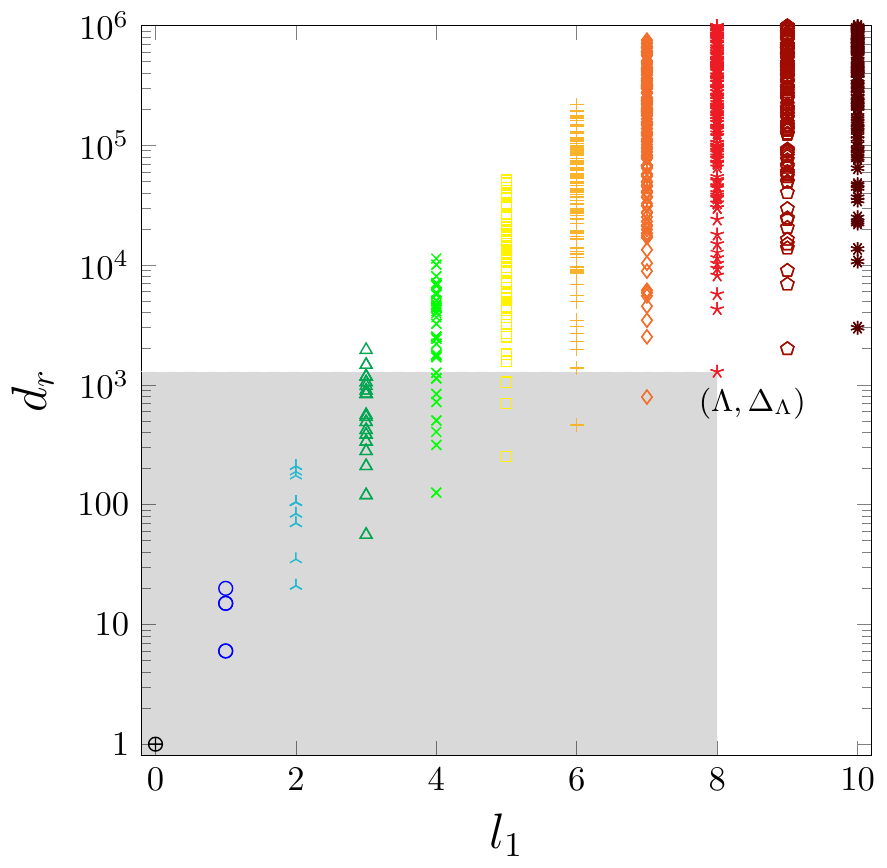}
\includegraphics[scale=0.85]{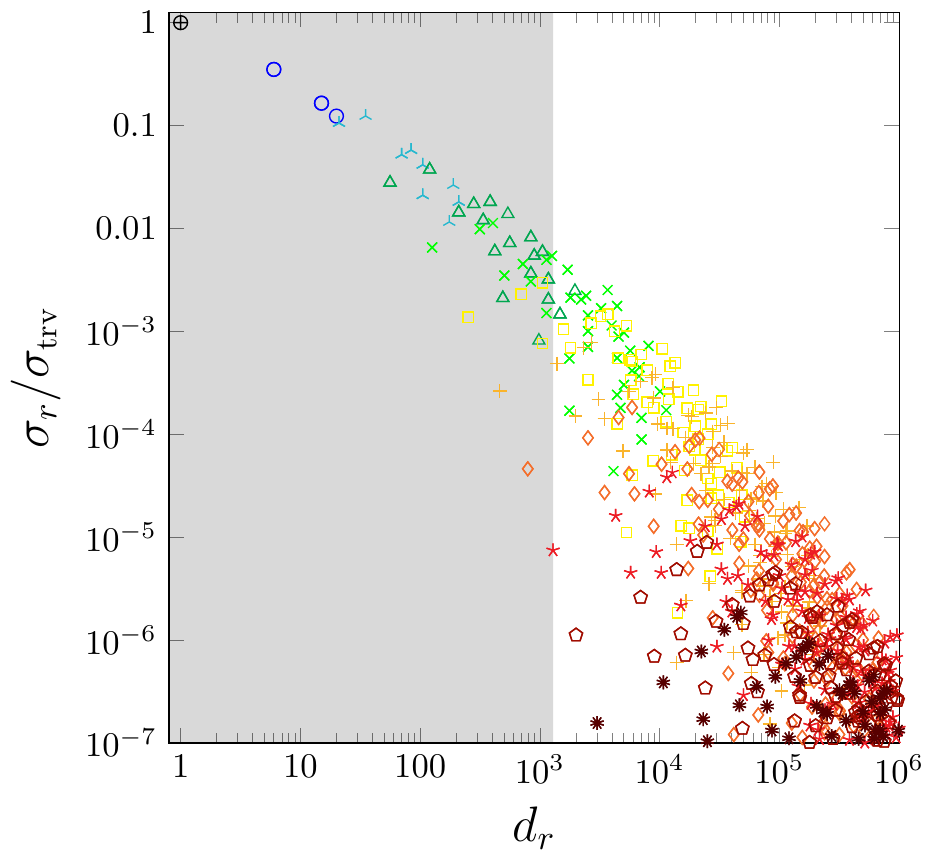}
\caption{(Left) The dimensionality
  of the representations of the ${\rm SU}(6)$
  group is
  plotted against $l_1$.
  The shaded region corresponds to $d_r\leq\Delta_\Lambda$
  and $l_1\leq\Lambda$ with $\Lambda=8$.
  (Right) The singular value $\sigma_r$ normalized by
  $\sigma_\text{trv}$
  is plotted
  against the dimensionality for the ${\rm SU}(6)$ gauge theory with $\lambda=3$.
  Different symbols correspond to different $l_1$ as shown
  in the left panel. The shaded region corresponds to $d_r\leq\Delta_\Lambda$
  with $\Lambda=8$.
}
\label{fig:DvSVandLvD}
\end{figure}

Another possibility is to put an upper bound on the
dimensionality \eqref{eq:representation_dimensionality} of the representation,
which is obviously the same for the complex conjugate pairs.
In Fig.~\ref{fig:DvSVandLvD} (Left), we plot the dimensionality $d_r$ of
each representation $r$ of the SU(6) group against $l_1$.
Let us note here that
the representation of the SU($N$) group that has the smallest dimensionality 
among those with the same $l_1=\Lambda$ is given by
the totally symmetric representation
and its conjugate, which
correspond to
$r=\{\Lambda,0, \cdots , 0\}$ and $\bar r=\{\Lambda, \cdots , \Lambda,0\}$,
respectively.
(See appendix \ref{sec:proof} for a proof.)
From \eqref{eq:representation_dimensionality},
one finds that the dimensionality of these representations is
\begin{align}
  \Delta_\Lambda&=
\prod_{j =2}^{N} \left(1+\frac{ \Lambda}{j-N}\right)
= {}_{\Lambda+N-1} {\rm C} _{\Lambda} \ ,
  \label{dimensionalitycutoff}
\end{align}
which grows monotonically with $\Lambda$.
Therefore, as a convenient way to put a cutoff on the dimensionality,
we use $d_r\leq\Delta_\Lambda$, which automatically implies
$l_1\leq\Lambda$ as is shown
by the shaded region in Fig.~\ref{fig:DvSVandLvD} (Left)
for $\Lambda=8$ in the SU(6) case.
This makes it easy to list all the representations below the cutoff.


In Fig.~\ref{fig:DvSVandLvD} (Right), we plot the singular value 
$\sigma_r=|\tilde\beta_r/d_r|$ of the fundamental tensor
for $\lambda=3$ corresponding to each representation $r$
against the dimensionality $d_r$.
Different symbols correspond to different $l_1$ as shown in the left panel.
We observe a clear tendency that
the singular value becomes small
as either $l_1$ or the dimensionality becomes large.

\begin{figure}
\centering
\includegraphics[scale=0.8]{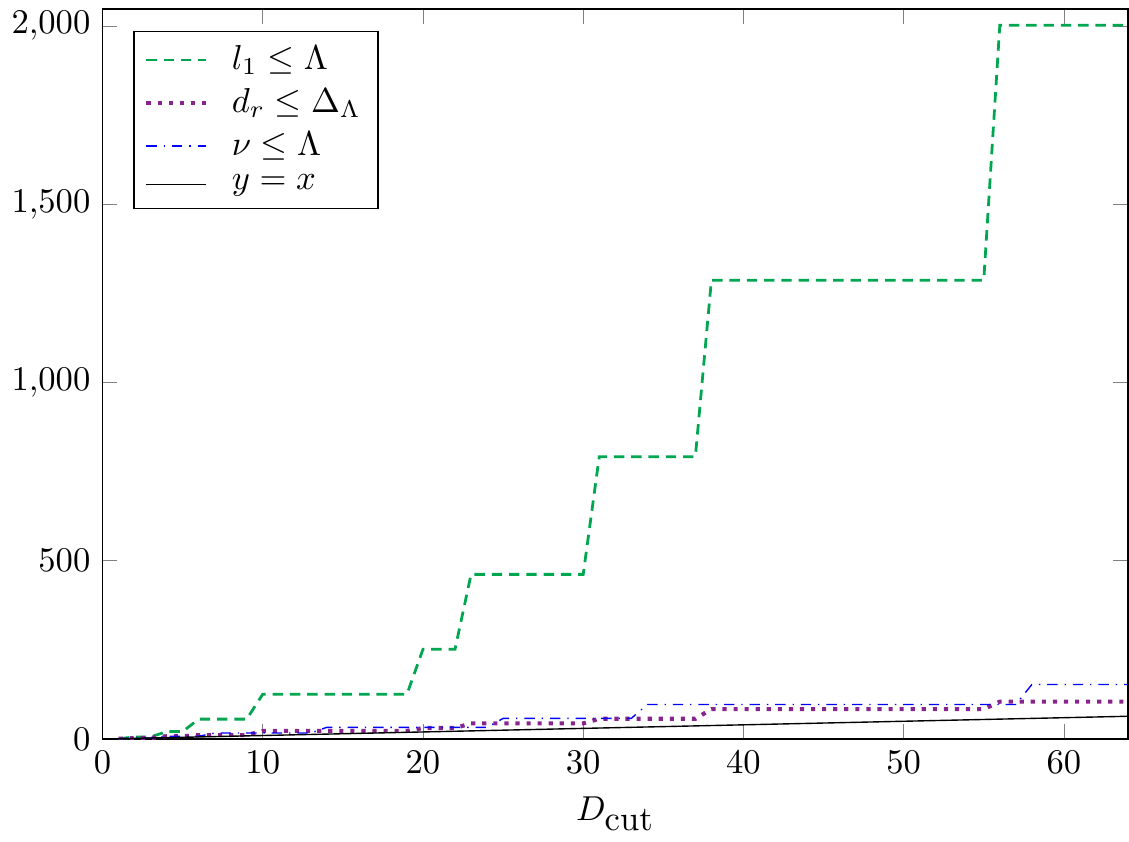}
\caption{The number of representations required
  for a given $D_\text{cut}$ is plotted against $D_\text{cut}$ for the
  ${\rm SU}(6)$ gauge theory with $\lambda=3$.
  The dashed, dotted and dash-dotted lines represent
  the results for the truncation by $l_1$, the dimensionality $d_r$,
  and $\nu \equiv \min_{q\in {\rm Z}} \sum_{i=1}^N |l_i+q|$,
  respectively. The solid line $y=x$ corresponds the theoretical lower
  bound.
}
\label{fig:comparison_of_cutoffs}
\end{figure}

In order to compare the efficiency of the two cutoff schemes,
we increase the cutoff in either cutoff scheme
until the largest $D_\text{cut}$ singular values of the fundamental tensor
do not change any more.
The number of representations below the cutoff
to achieve a given value of $D_\text{cut}$
is plotted in Fig.~\ref{fig:comparison_of_cutoffs}.
It is clear that the number of representations
is significantly smaller for the cutoff scheme with the dimensionality.
It is also worth noting that
in the case of SU(2),
the two cutoff schemes reduce to the truncation by spins $S$ since
$l_1=2S$ and $d_r=2S + 1$.
What we find here is that for $N>2$, it is far more efficient
to truncate the representations by the dimensionality $d_r$ than by $l_1$.


Next we consider the $\G{U}(N)$ case using the
relationship \eqref{eq:shifting_by_q}
to the representations of $\G{SU}(N)$,
which can be truncated in the way described above.
The remaining task is to introduce a cutoff on $q$ in \eqref{eq:shifting_by_q}.
Note that the complex conjugate representation of $\{ l_i ' \}$ is given
by $\{ \bar{l}_i ' \}$, where $\bar{l}_i  ' = - l_{N+1-i} '$.
Therefore, as a simple cutoff on $q$ that does not discriminate the complex conjugates,
we
impose
\begin{align}
  \max_i |l_i '|
  =   \left|q + \frac{l_1}{2}\right| + \frac{l_1}{2}\leq \Lambda_q  \ ,
\end{align}
where $\Lambda_q$ is an integer representing the cutoff on $q$.
Note that this is a natural generalization of the U(1) case,
where we truncate the representations by the charge.

As yet another way of truncation,
we can think of $\sum_{i=1}^N |l_i '| \le \Lambda$
for the U($N$) case, which is clearly invariant under
complex conjugation.
The SU($N$) analogue of this truncation is given by
$\nu \equiv \min_{q\in {\rm Z}} \sum_{i=1}^N |l_i+q| \le \Lambda$.
The efficiency of this truncation scheme turns out to be
comparable to
the one with the dimensionality
as we can see from Fig.~\ref{fig:comparison_of_cutoffs}.


\subsection{properties of the singular-value spectrum}
\label{subsection:dependence-coupling}

\begin{figure}[t]
\centering
\includegraphics[scale=0.8]{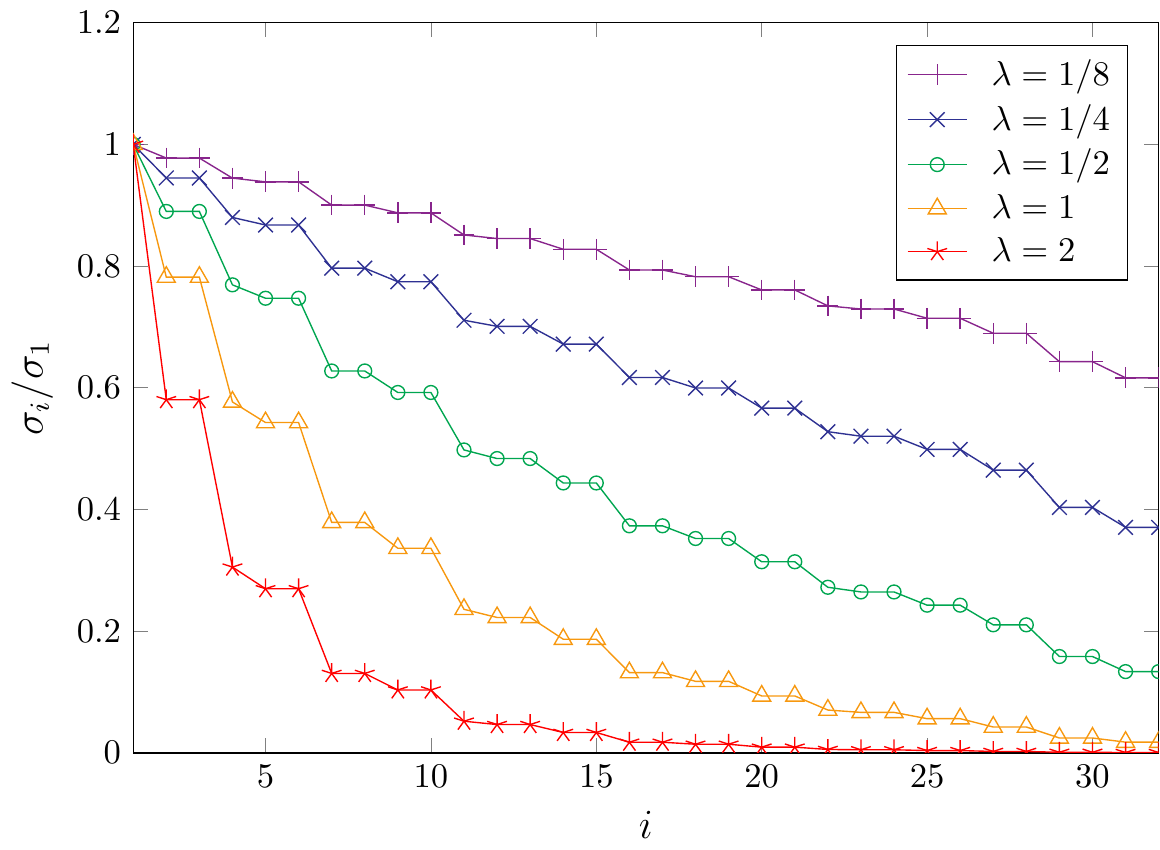}

\caption{The singular values $\sigma_i/\sigma_1$ ($i=1,2, \cdots ,32$) of the
  fundamental tensor normalized by the largest one $\sigma_1$
  are plotted in the descending order
  for the $\G{SU}(3)$ gauge theory with various values of the coupling constant.}
\label{fig:SVvsLambda}
\end{figure}

Using the strategy for restricting the number of representations
proposed in the previous section,
we can obtain the singular-value spectrum correctly up to any given $D_\text{cut}$.
Below we discuss various properties of the singular-value spectrum thus obtained.
First we discuss how the singular-value spectrum depends on the
coupling constant.
In Fig.~\ref{fig:SVvsLambda}, we plot the singular values $\sigma_i$ ($i=1 , 2, \cdots$)
sorted in the descending order
for the SU(3) case with various values of the coupling constant.
The singular values are normalized by the
largest one $\sigma_1$, which corresponds to the trivial representation
as we mentioned at the end of Section \ref{sec:sing-val-2D}.
We find that the singular-value spectrum falls off more slowly
at weak coupling, where the correlation length becomes large.



Next we discuss the behavior of the singular values
at large $N$.
%
%
In particular, we will see
that the ratio of the singular values $\sigma_i/\sigma_1$
becomes finite in the large-$N$ limit.
We also find, in the U($N$) case, that the large-$N$ behavior
of the singular values changes qualitatively at the
critical coupling of the Gross-Witten-Wadia phase transition.

\begin{figure}[t]
\centering
\includegraphics[scale=0.8]{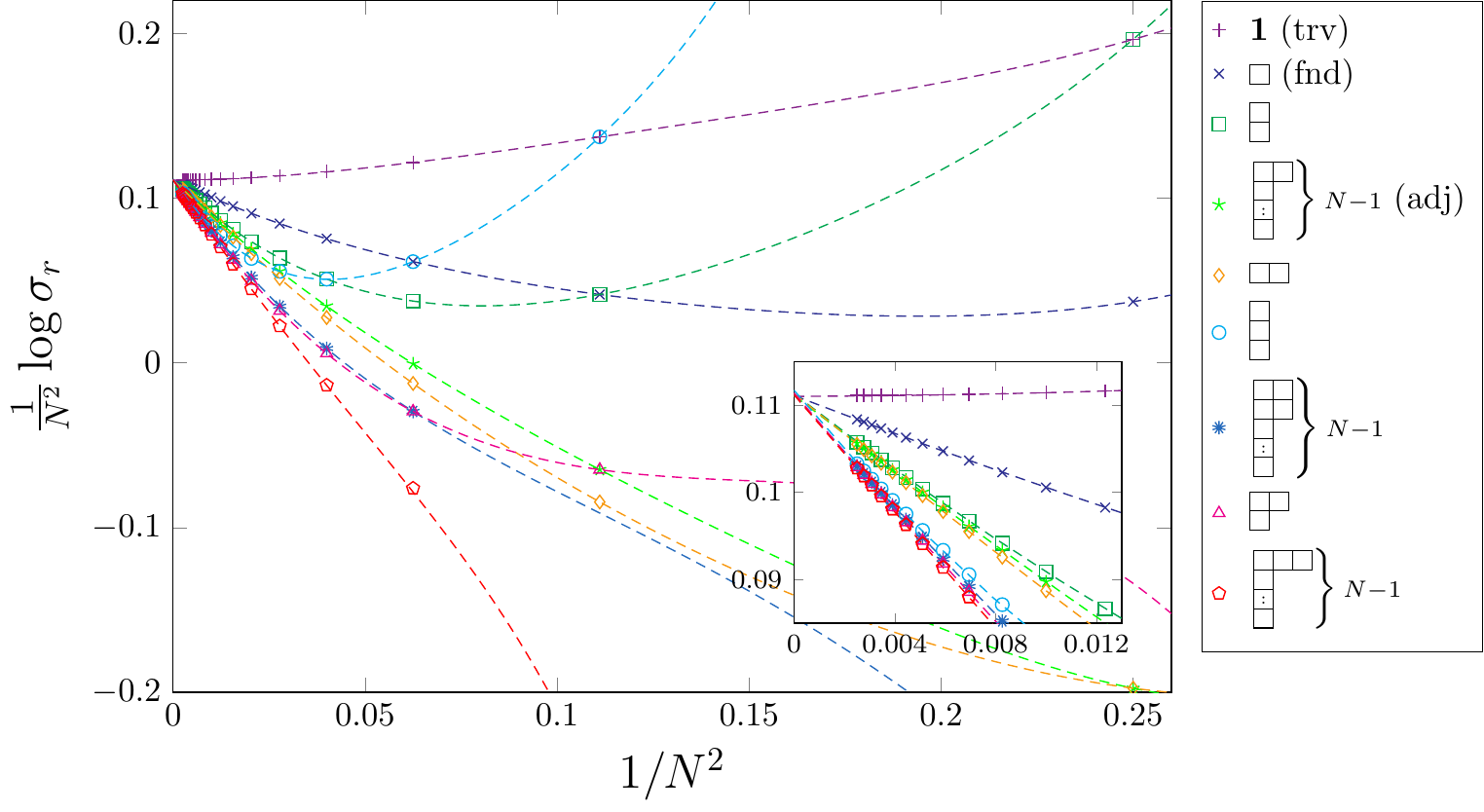}
\caption{The singular values $\frac{1}{N^2}\log\sigma_r$
  are plotted against $1/N^2$ for the $\G{SU}(N)$ gauge theory with $\lambda=3$.
  Different symbols correspond to different sequences of representations
  as shown on the right of the plot. 
  Each point is actually doubly degenerate
  except for real representations since
  representations that are complex conjugate to each other give identical singular values.
  We restrict ourselves to the sequence of representations that
  correspond to the 16 largest singular values
  in the large-$N$ limit.
  The dashed lines represent fits to a quartic polynomial in $1/N^2$
  for each sequence using $N \ge 2$,
  which turn out to merge at
   $C^{(0)} =\frac{1}{9}$ in the large-$N$ limit.
}
\label{fig:SVvsN2}
\end{figure}

\begin{figure}
\centering
\includegraphics[scale=0.8]{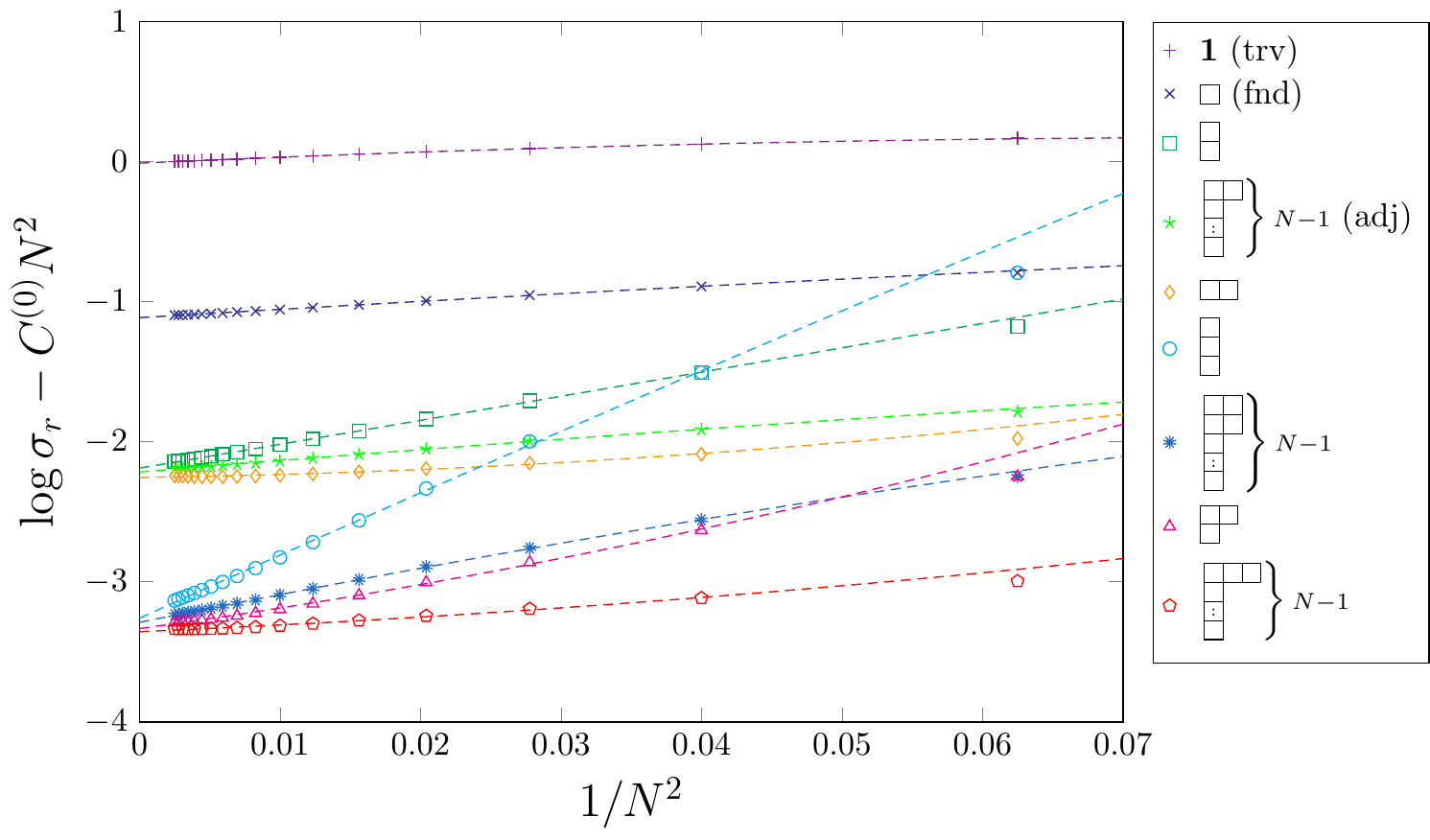}
\includegraphics[scale=0.8]{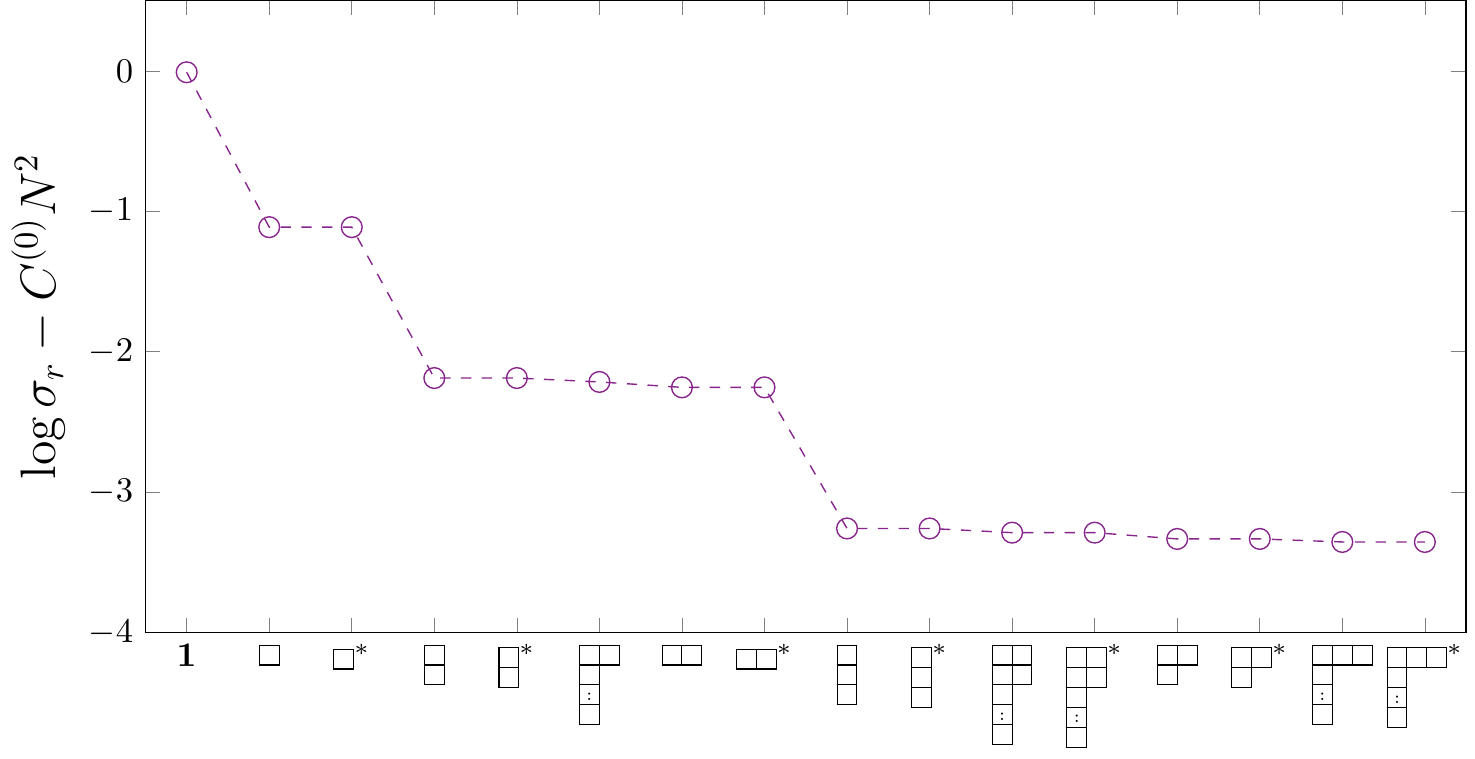}
\caption{(Top) The quantity $(\log  \sigma_r  - C^{(0)} N^2)$
  is plotted against $1/N^2$
  for the $\G{SU}(N)$ gauge theory with $\lambda=3$.
  Different symbols correspond to different sequences of representations as shown
  on the right.
  Each point is actually doubly degenerate
  except for real representations since
  representations that are complex conjugate to each other give identical singular values.
  The dashed lines represent fits to a quadratic polynomial in $1/N^2$ for each sequence using $N\ge 5$.
  (Bottom) The large-$N$ limit of
$(\log  \sigma_r  - C^{(0)} N^2)$
  obtained from
  the top panel is shown for each sequence of representations
  in the descending order.
}
\label{fig:largeN-convergence}
\end{figure}

In order to discuss the large-$N$ behavior of the singular values,
we define a sequence of representations for increasing $N$.
For example,
the trivial, (anti-)fundamental and adjoint representations,
which are defined for
$N=2,3, \cdots $
can be regarded as sequences of representations.
Generalizing these examples, we define
a sequence of representations,
in the U($N$) case, by specifying
$n\ge 0$ positive entries and $m \ge 0$ negative entries of $\{ l'_i \}$ as
\begin{align}
r^{({\rm U})} =
\{ l'_1 , \cdots , l_n' , 0 , \cdots , 0 , l'_{N-m+1} , \cdots , l'_N \} \ .
\label{sequence-UN}
\end{align}
As $N$ increases, the number of zero entries in the middle
increases while keeping the structure at both ends of $\{ l'_i \} $ fixed.
By adding charge $q\in {\bf Z}$
to the representations \eqref{sequence-UN},
we can obtain other sequences
\begin{align}
(r^{({\rm U})},q)  \equiv
\{ l'_1 + q , \cdots , l_n'+q , q , \cdots , q ,
l'_{N-m+1}+q , \cdots , l'_N+q \} \ .
\label{sequence-UN-q}
\end{align}
In the SU($N$) case, we can define
a sequence of representations
$r^{({\rm SU})}\equiv \{ l_i \} $,
where $l_i = l'_i-l'_N$ ($i=1, \cdots , N$), 
corresponding to \eqref{sequence-UN}.
In what follows, we discuss the large-$N$ behavior of the
singular values for the sequences of representations thus defined.



Let us start with the $\G{SU}(N)$ gauge theories.
We observe that the singular-value spectrum has a large-$N$
behavior
consistent with
\begin{align}
  \log\sigma_r &= \sum_{k=0}^{\infty} C_r ^{(k)} N^{2(1-k)}
  = C_r ^{(0)} N^2 + C_r ^{(1)}  + C_r ^{(2)} N^{-2}  + \cdots   \ ,
  \label{eq:large-N singular values}
\end{align}
where $r$ corresponds to a sequence of representations for increasing $N$
and $C_r^{(k)}$ depend on the coupling constant $\lambda$.
In Fig.~\ref{fig:SVvsN2} we plot $\frac{1}{N^2}\log\sigma_r$ against
$\frac{1}{N^2}$ for the $\G{SU}(N)$ gauge theory\footnote{We have
  obtained qualitatively the same behavior at $\lambda=1.5$, which is
  below the Gross-Witten-Wadia phase transition point $\lambda=2$,
  in contrast to the situation in the U($N$) case discussed below.}
with $\lambda=3$.
All the sequences listed here, which correspond to 16 largest singular values
in the large-$N$ limit, can be fitted 
by \eqref{eq:large-N singular values} including terms up to $k=4$.
It is remarkable that the fit works nicely including the data points for $N=2$.
In fact, we find that
the large-$N$ limit
$C_r^{(0)} = \lim_{N \rightarrow \infty }\frac{1}{N^2}\log\sigma_r$
for any sequence of representations
is universal, and
it is given by $C^{(0)}$ defined in \eqref{eq:GWW_vacuum_energy}.
Thus we conclude that
the ratio of the singular values $\sigma_r /\sigma_{\text{trv}}$
is finite in the large-$N$ limit.
In other words, the singular-value spectrum has a
definite profile given by $C_r^{(1)}$ in the large-$N$ limit
up to an overall factor.

We can obtain the large-$N$ limit of the singular-value spectrum
explicitly as follows.
In Fig.~\ref{fig:largeN-convergence} (Top)
we plot
$(\log  \sigma_r  - C^{(0)} N^2)$
against $1/N^2$
for the $\G{SU}(N)$ gauge theory with $\lambda=3$.
Fitting
the data for $N \ge 5$
to a quadratic polynomial for each sequence,
we can make an extrapolation to $N=\infty$.
In Fig.~\ref{fig:largeN-convergence} (Bottom)
we plot the
large-$N$ limit
of
$(\log  \sigma_r  - C^{(0)} N^2)$
thus obtained
for each sequence of representations in the descending order.

\begin{figure}
\centering
\includegraphics[scale=0.8]{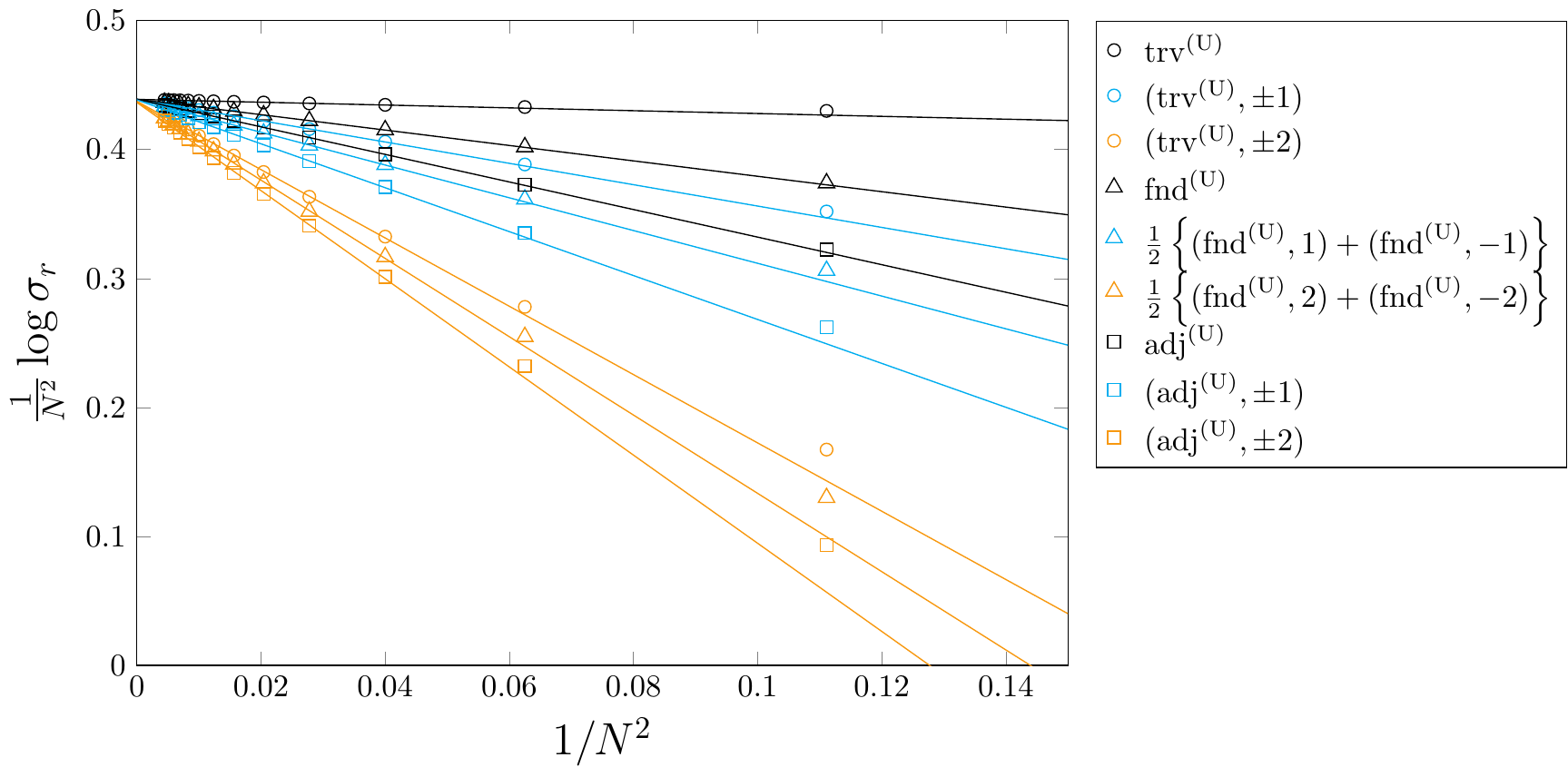}
\includegraphics[scale=0.8]{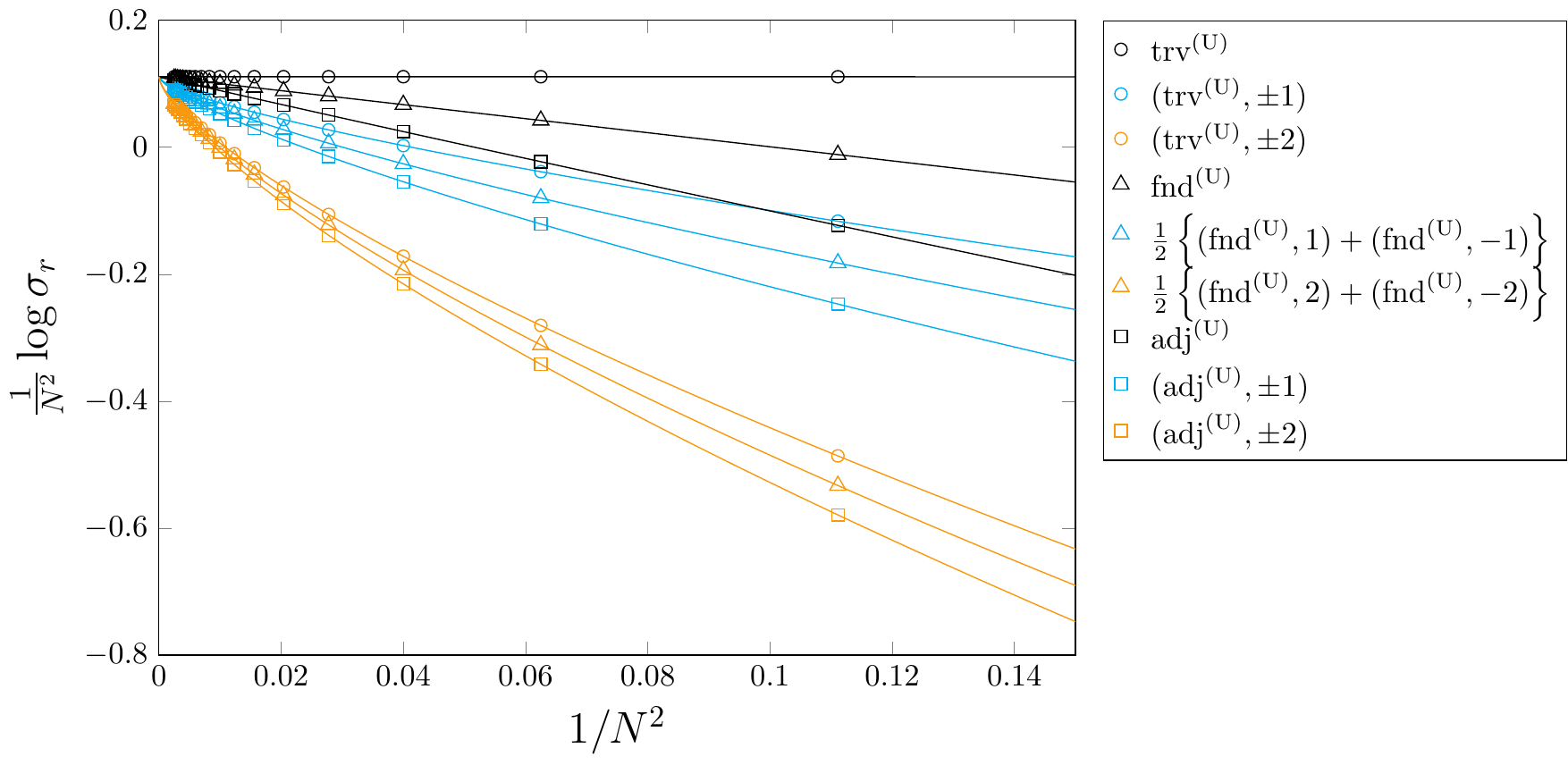}
\caption{(Top) The singular values $\frac{1}{N^2}\log\sigma_r$
for the trivial (black circles), fundamental (black triangles)
  and adjoint (black squares) representations of
  $\G{U}(N)$
  are plotted against $1/N^2$ for the $\G{U}(N)$ gauge theory with
  $\lambda=1.5$.
  We also add some charge to these representations
  and plot the corresponding
  singular values in the same figure with different colors.
  The lines represent fits of the results for each representation
  with $N\ge 7$ to
  $C + c/N^2$ for $\lambda=1.5$.
  (Bottom) A similar plot for $\lambda=3$.
  The lines represent fits of the results for each representation
  with $N\ge 3$ to
  $C + a/N + b \chi_N/N^2 + c/N^2$.
  The fits for the sequence of representations without charge
  yield $a=b=0$ within fitting errors.
%
}
\label{fig:SVvsN2UN}
\end{figure}

Next let us consider the $\G{U}(N)$ case.
The large-$N$ behavior of the singular-value spectrum
changes qualitatively as one crosses the critical point $\lambda=2$
of the Gross-Witten-Wadia phase transition \eqref{eq:GWW_vacuum_energy}.
In Fig.~\ref{fig:SVvsN2UN},
we plot the singular values $\frac{1}{N^2}\log\sigma_r$
for the trivial (black circles), fundamental (black triangles) and
adjoint representations (black squares) of $\G{U}(N)$
against $1/N^2$
in the $\G{U}(N)$ gauge theory
with $\lambda=1.5$ (Top) and $\lambda=3$ (Bottom).
  We also add some charge to these representations
  and plot the corresponding
  singular values in the same figure with different
  colors\footnote{Since the fundamental representation is not real
    unlike the trivial and adjoint representations, the representations
    one gets by adding charge $\pm q$ to it are not complex conjugate
    to each other and hence the corresponding singular values acquires
    $1/N$ terms with opposite signs as we see
    in Section \ref{sec:SV-largeN-theta}.
    Here we take an average of the results for the representations
    with charge $\pm q$ to cancel the $1/N$ terms
    for the sake of simplicity.}.
  In both cases, the large-$N$ limit of $\frac{1}{N^2}\log\sigma_r$
  is common to all representations and agree with \eqref{eq:GWW_vacuum_energy},
  but the way the limit is approached turns out to be
  qualitatively different for 
  $\lambda=1.5$ and $\lambda=3$.

  In the weak coupling phase $\lambda<2$,
  we find that the singular values behave as\footnote{The $1/N$
    terms appear for the complex representations with nonzero charge.}
\begin{align}
  \log\sigma_r &= C^{(0)} N^2 + C_r ^{(1)} + {\rm O}\left(\frac{1}{N}\right)  \ .
  \label{eq:large-N singular values-UN}
\end{align}
The coefficient $C^{(0)}$ of the leading term is common to all representations
and is given by \eqref{eq:GWW_vacuum_energy},
while the coefficient $C_r^{(1)}$ for the sub-leading term
depends on the
sequence of representations.
This is confirmed in Fig.~\ref{fig:SVvsN2UN} (Top).
Thus,
the singular-value spectrum has a
definite profile given by $C_r^{(1)}$,
which is different from that in the $\G{SU}(N)$ case.

\begin{figure}
\centering
\includegraphics[scale=0.8]{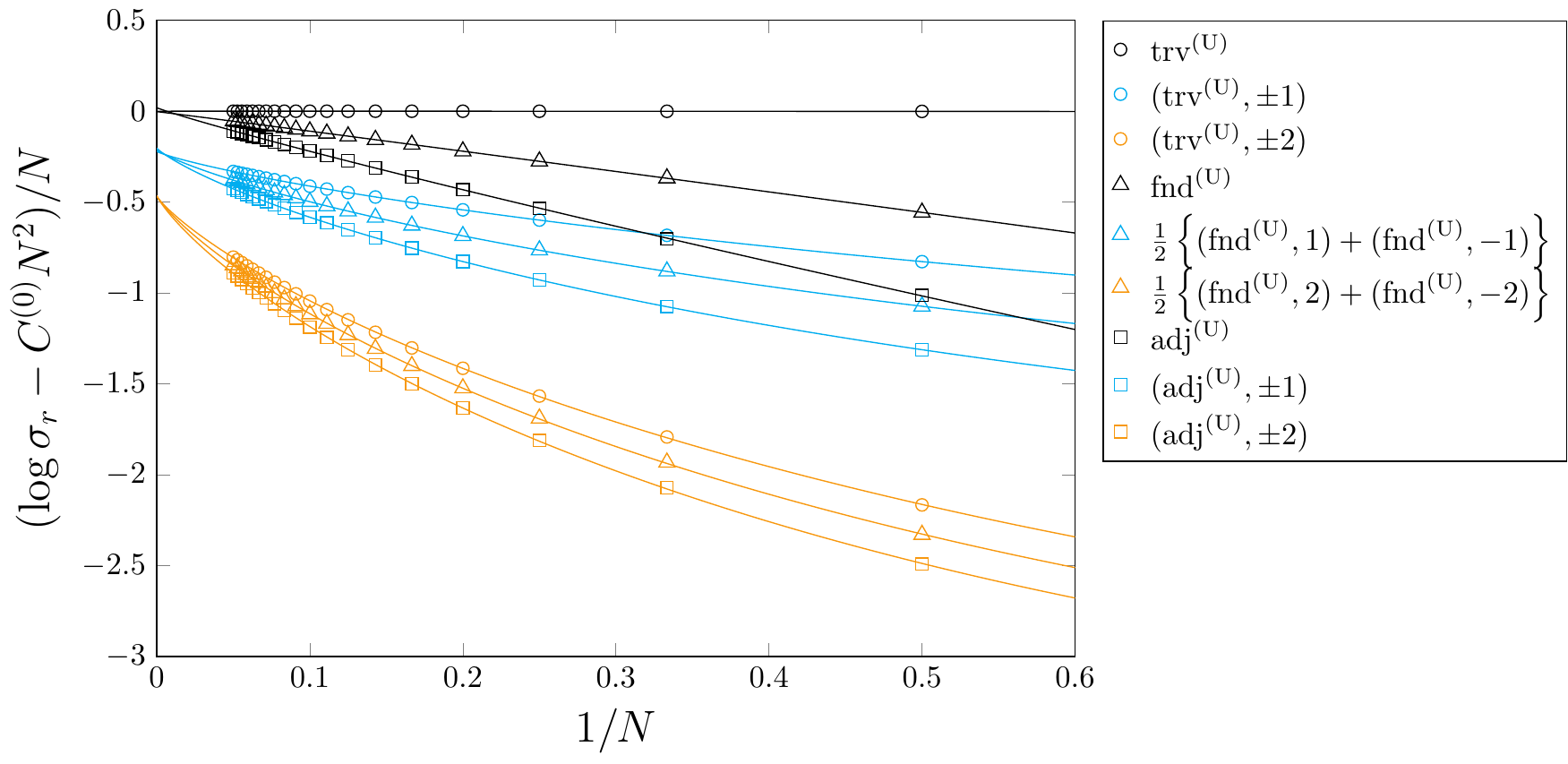}
\caption{The quantity
    $(\log\sigma_r - C^{(0)} N^2)/N$
  is plotted against $1/N$
  for the trivial, fundamental and adjoint representations
  in the $\G{U}(N)$ gauge theory with $\lambda=3$.
  We also plot the same quantity for representations
  with nonzero charge
  with different colors.
  The lines represent fits
  of the results for each representation with $N\ge 2$
  to $a  + b \chi_N/N + c/N$.
  For the sequences of representations without charge,
  the fits yield $a=b=0$ within fitting errors.
  %
}
\label{fig:SV-subtract}
\end{figure}

In the strong coupling phase $\lambda>2$,
the large-$N$ behavior of the singular-value spectrum
turns out to be more complicated.
We find that
the singular values behave as
\eqref{eq:large-N singular values-UN}
for the sequence \eqref{sequence-UN} of representations
without charge.
For sequences of representations
\eqref{sequence-UN-q} with nonzero charge,
the singular values behave as
\begin{align}
  \log\sigma_r &= C^{(0)} N^2 + C_r ^{(1)} 
  + a N + b\chi_N +   {\rm O}\left(\frac{1}{N}\right)
  \label{eq:large-N singular values-UN-SC}
\end{align}
at large $N$,
where the definition of $\chi_N$ is given in \eqref{eq:susceptibility_predictions},
which has an O($\log N$) term for $\lambda >2$ that we are considering.
Note that \eqref{eq:large-N singular values-UN-SC} involves
an O($N$) term, whose coefficient $a$ turns out to be negative.
This implies
that the corresponding singular values are exponentially suppressed
as $\sigma_r /\sigma_\text{trv} \sim e^{-|a|N}$ at large $N$,
and hence disappear from
the singular-value spectrum in the large-$N$ limit.
These behaviors are confirmed
in Fig.~\ref{fig:SVvsN2UN} (Bottom).

The coefficient of the O($N$) term is actually the same
for the same charge,
as can be seen from Fig.~\ref{fig:SV-subtract},
where we plot $(\log \sigma_r - C^{(0)}N^2)/N$ against $1/N$.
We find that the curves merge at the same point at $N=\infty$
for the same charge.
A theoretical understanding of the ${\rm O}(N)$ term
as well as the term involving $\chi_N$
in \eqref{eq:large-N singular values-UN-SC}
is provided in Section \ref{sec:SV-largeN-theta}.

%

Note also that \eqref{eq:character_coefficient} implies
\begin{equation}
  \sigma_{r^\text{(SU)}}
  =\sum_{q=-\infty}^\infty  \sigma_{(r^\text{(U)},q)} \ .
    \label{eq:LargeN_neutralrep}
\end{equation}
This has an interesting implication for $\lambda >2$
since in that case the sum on the right-hand side is dominated
at large $N$
by a single term
with $q=0$.
Namely, the ratio of the singular value of $(r^{\rm (U)},0)$
to that of the corresponding $r^{\rm (SU)}$ becomes unity in the large-$N$ limit.
This means that
the coefficient
$C_{r}^{(1)}$ in \eqref{eq:large-N singular values}
and \eqref{eq:large-N singular values-UN}
representing the sub-leading term
is common to
the $\G{U}(N)$ and $\G{SU}(N)$ gauge theories for the corresponding representations.
Thus, for $\lambda > 2$,
the singular-value spectrum
for the U($N$) gauge theory
not only has a definite profile in the large-$N$ limit
but also
agrees with the profile for the SU($N$) gauge theory.

\subsection{a novel interpretation of the Eguchi-Kawai reduction}
\label{subsection:EK-equiv}

In this section, we discuss the implications of the properties of the singular-value
spectrum found in the previous section.
In particular, we provide a new interpretation of
the Eguchi-Kawai reduction \cite{Eguchi1982},
which states the volume independence in the large-$N$ limit.

As we have seen in the previous section,
the leading large-$N$ behavior of the singular values $\sigma_r$
is given by $e^{C^{(0)} N^2}$ independently of the sequence of representations
for both SU($N$) and U($N$) gauge theories,
where $C^{(0)}$ is given by \eqref{eq:GWW_vacuum_energy}.
Factoring out this common factor, one obtains a finite large-$N$ limit
\begin{align}
  \lim_{N\rightarrow \infty} e^{-C^{(0)} N^2} \sigma_r = e^{C_r^{(1)}} \ ,
    \label{eq:sigma-infinite-N}
\end{align}
which depends on the sequence of representations $r$.
In the U($N$) gauge theory at $\lambda > 2$, the limit in 
\eqref{eq:sigma-infinite-N} becomes nonzero only for sequences
of representations without charge,
which agrees with the limit for the corresponding
sequences of representations in the SU($N$) gauge theory.
Using \eqref{eq:sigma-infinite-N} in
\eqref{eq:TRGpartition}, we find that
the free energy density is given by
\begin{equation}
 F \equiv   \frac{1}{N^2 V} \log Z = C^{(0)} +
  \frac{1}{VN^2}\log \left( \sum_r  e^{VC_r^{(1)}} \right)
  + \cdots \ ,
  \label{eq:EK_volume_independent}
\end{equation}
which implies that
the free energy density
becomes volume independent in the large-$N$ limit,
and it is given by \eqref{eq:GWW_vacuum_energy}.
This is nothing but the Eguchi-Kawai reduction \cite{Eguchi1982}.

%

\begin{figure}
\centering
\includegraphics[scale=0.75]{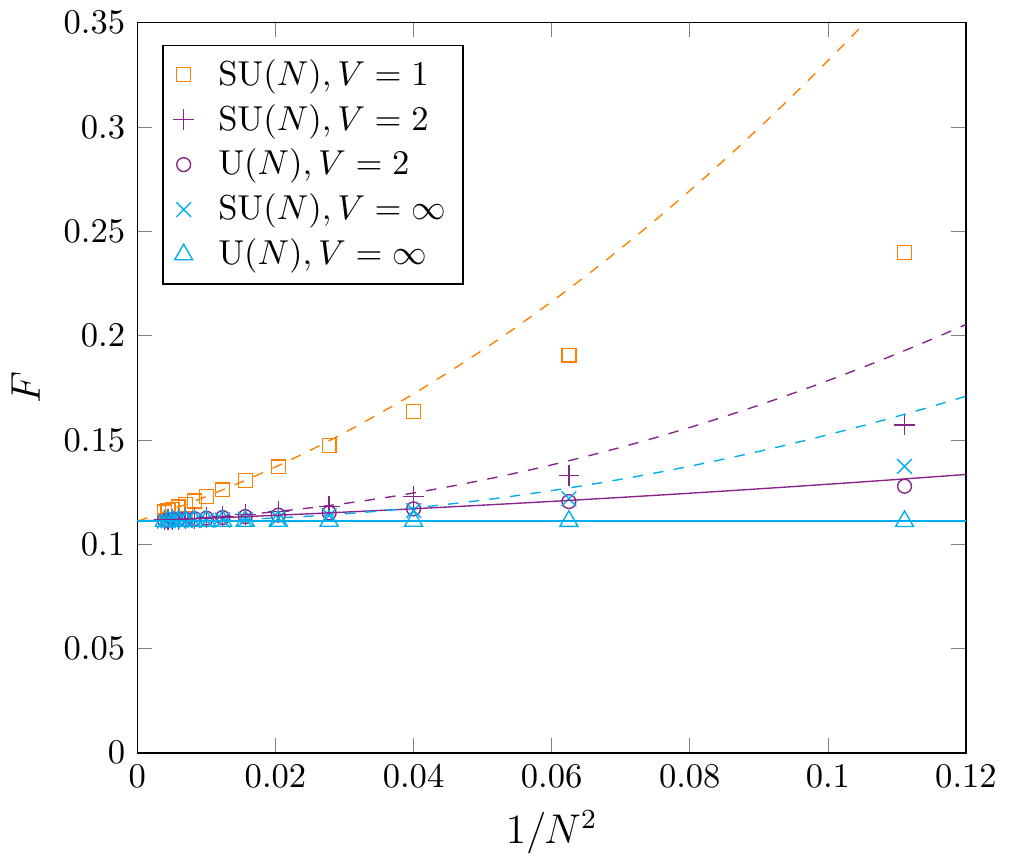}
\includegraphics[scale=0.75]{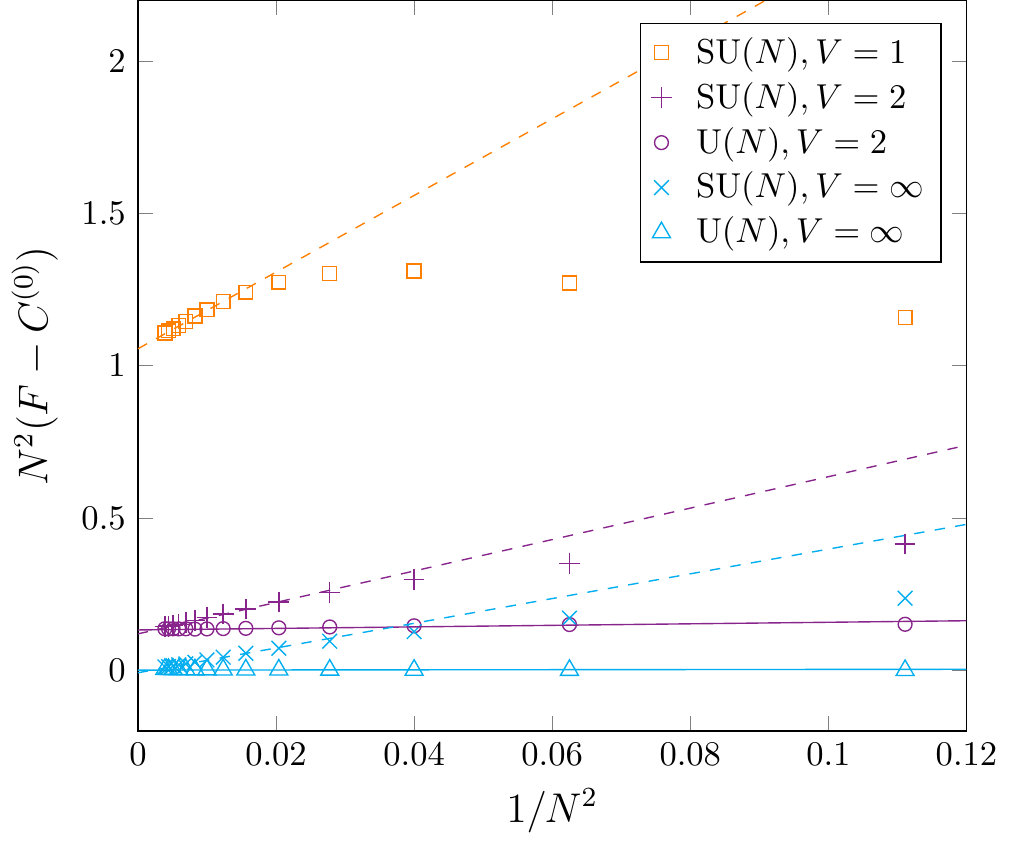}
\caption{(Left) The free energy density
  $F$
  is plotted against $1/N^2$ for the U($N$) and SU($N$) gauge theories
  at $\lambda=3$
  with $V=1$, $2$ and $\infty$
  using the bond dimension $D_\text{cut}=64$.
  In the case of $V=1$, there is no distinction between U($N$) and SU($N$) gauge theories,
  so we plot only the results for the latter.
  The solid and dashed lines are fits to the function $a + b/N^2+c/N^4$
  for the U($N$) and SU($N$) gauge theories, respectively.
  (Right) The quantity $N^2(F-C^{(0)})$ is plotted against $1/N^2$
  for the U($N$) and SU($N$) gauge theories at $\lambda=3$
  with $V=1$, $2$ and $\infty$ using the bond dimension $D_\text{cut}=64$.
  The solid and dashed lines are fits to the function $a + b/N^2$
  for the U($N$) and SU($N$) gauge theories, respectively.
  The extrapolated values at $N=\infty$ give the coefficients
  of the O($1/N^2$) terms in the free energy density $F$.
}
\label{fig:EK}
\end{figure}

In Fig.~\ref{fig:EK} (Left), we plot the free energy density $F$
against $1/N^2$ for the U($N$) and SU($N$) gauge theories
at $\lambda=3$ 
with $V=1$, $2$ and $\infty$
using the bond dimension $D_\text{cut}=64$,
which is large enough as we see
in Section \ref{subsection:DcutDependence}.
The results for $V=\infty$ converge
to \eqref{eq:GWW_vacuum_energy} in the large-$N$ limit
as it should\footnote{In the case of U($N$) gauge theory at $V=\infty$,
  we find that the free energy density is actually independent of $N$,
  which is due to the fact that the singular value for the trivial representation
  is given by
  $\sigma_{\rm trv}= C^{(0)}N^2$
  for arbitrary $N$
  in the strong coupling phase
  $\lambda>2$ as one can see from
  Fig.~\ref{fig:SVvsN2UN} (Bottom).}, 
but we also observe that the results for $V=1$ and $2$ converge to
the same value, which confirms the Eguchi-Kawai reduction.

In fact,
  we find that the results for $V=1$ are the same for
  the U($N$) and SU($N$) gauge theories even at finite $N$
  as one can see by setting $V=1$ in \eqref{eq:TRGpartition} and using \eqref{eq:LargeN_neutralrep}.
  This can be understood from the fact
  that the two theories are equivalent at $V=1$
  since the U(1) part of the gauge field cancels in the plaquette \eqref{def-plaquette}.
We also notice from our argument above that at $\lambda > 2$,
the O($1/N^2$) term in \eqref{eq:EK_volume_independent}
should agree for the U($N$) and SU($N$) gauge theories.
This is confirmed in Fig.~\ref{fig:EK} (Right) for $V=2$ by extrapolating
$N^2(F-C^{(0)})$ to $N=\infty$, which gives identical values for 
the U($N$) and SU($N$) gauge theories. Note that
this statement does not hold at $\lambda < 2$.


Thus, the Eguchi-Kawai reduction in the present model
results from
the simple $V$-dependence of the partition function
\eqref{eq:TRGpartition} and the large-$N$ behavior
\eqref{eq:sigma-infinite-N}
of the singular values
with the common $C^{(0)}$.
The simple $V$-dependence \eqref{eq:TRGpartition}
may also be viewed as a consequence
of the fact that the singular-value spectrum of the fundamental tensor
has certain self-similarity under the coarse-graining procedure
as seen in \eqref{eq:coarse-grained-tensor}
and \eqref{eq:coarse-grained-tensor-final}.

In $d \ge 3$ dimensional cases, the volume independence in the $N \rightarrow \infty$ limit
is expected only for $V > V_{\rm cr}$
since the SSB of $U(1)^{d}$ symmetry \cite{Bhanot:1982sh}
occurs for $V \le  V_{\rm cr}$
\cite{Narayanan:2003fc,Kiskis:2003rd},
which invalidates the proof of the Eguchi-Kawai reduction \cite{Eguchi1982}.
Note also that the partition function does not take a simple form like
\eqref{eq:TRGpartition} any more.
However, we speculate that the volume independence for $V > V_{\rm cr}$
may be understood by some sort of self-similarity
that manifests itself after some steps of coarse graining at large $N$.


\section{Explicit results with a bond dimension $D_\text{cut}$}
\label{sec:explicit_results}

Given the singular values of the fundamental tensor
up to some bond dimension $D_\text{cut}$,
we can obtain explicit results
for various observables.
We discuss how the results depend on $D_\text{cut}$,
and show, in particular, that the finite $D_\text{cut}$ effects
become severe for small $N$, small volume and at weak coupling.
We can obtain explicit results even in such cases.
As an example, we show how the Gross-Witten-Wadia phase transition appears
as $N$ increases.

\subsection{the \texorpdfstring{$D_\text{cut}$}{Dcut}-dependence}
\label{subsection:DcutDependence}

Let us first discuss the $D_\text{cut}$-dependence.
Since the observables are typically obtained by taking
the derivatives of the free energy density $F$
\eqref{eq:EK_volume_independent} in the TRG,
let us consider the $D_\text{cut}$-dependence of $F$.
Note first that the $D_\text{cut}$-dependence disappears in the
infinite-volume limit
since the trivial representation dominates in that limit
as we mentioned at the end of Section \ref{sec:sing-val-2D}.
Therefore, we focus on
the free energy density $\tilde{F}(D_\text{cut})$ obtained for finite $D_\text{cut}$
with $V=1$, where finite $D_\text{cut}$ effects become the severest.
%

\begin{figure}
\centering
\includegraphics[scale=0.8]{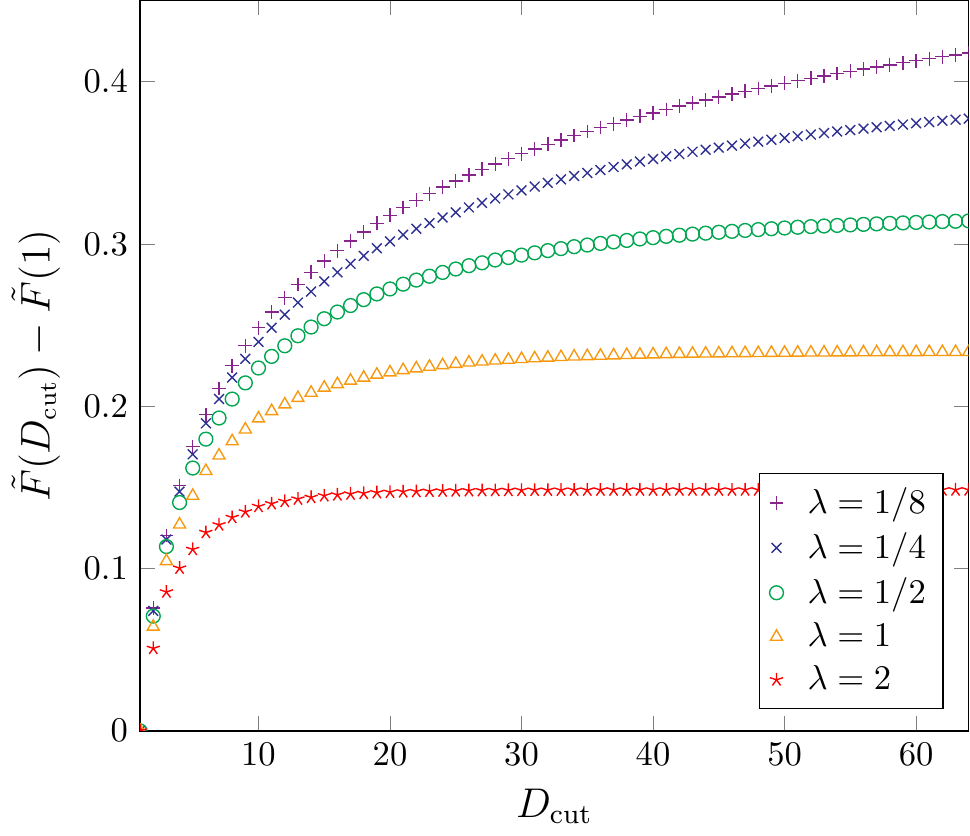}
\includegraphics[scale=0.8]{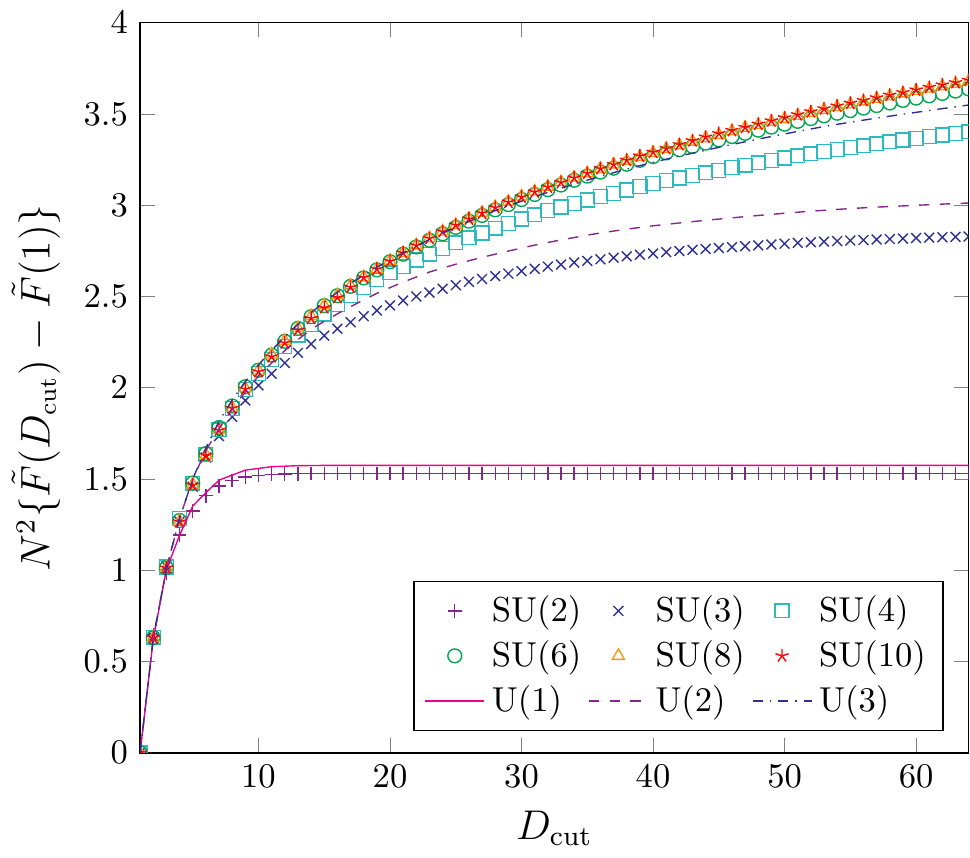}
\caption{(Left) The free energy density
  $\Big\{ \tilde{F}(D_\text{cut}) - \tilde{F}(1) \Big\}$
  for $V=1$ is plotted against $D_\text{cut}$
  for the $\G{SU}(3)$ gauge theory
  with various values of the coupling constant, which
  are used also in Fig.~\ref{fig:SVvsLambda}.
  (Right) The free energy density
  $N^2 \Big\{ \tilde{F}(D_\text{cut}) - \tilde{F}(1) \Big\}$
  for $V=1$
  is plotted with different normalization against $D_\text{cut}$
  for the U($N$) and SU($N$) gauge theories
  with various $N$ at $\lambda=0.5$.}
\label{fig:DcutDependence}
\end{figure}

In Fig.~\ref{fig:DcutDependence} (Left),
we plot the free energy density $\Big\{ \tilde{F}(D_\text{cut}) - \tilde{F}(1) \Big\}$
against $D_\text{cut}$
for the $\G{SU}(3)$ gauge theory with various values of the coupling constant, which
are used also in Fig.~\ref{fig:SVvsLambda}.
We find that the free energy density converges faster at stronger coupling.
This is consistent with the singular-value distribution
in Fig.~\ref{fig:SVvsLambda},
which falls off faster at larger $\lambda$.

\mycomment{
    \begin{figure}
    \centering
    \includegraphics[scale=1]{figures/DcutOneOnly.pdf}
    \caption{The difference between the free energy density at $V=1$ obtained
      with $D_\text{cut}=1$ and $D_\text{cut}=\infty$ (extrapolated)
      as a function of $1/N^2$.
      The finite-$D_\text{cut}$ effect vanishes at large-$N$ as $1/N^2$,
      which interestingly suggests that using $D_\text{cut}=1$ is sufficient at large enough $N$.}
    \label{fig:DcutOneOnly}
    \end{figure}
}


In fact, the finite $D_\text{cut}$ effects
vanish as ${\rm O}(1/N^2)$
at large $N$
since finite volume effects,
which cause the finite $D_\text{cut}$ effects,
are suppressed by $1/N^2$ due to the Eguchi-Kawai reduction.
In order to see this more explicitly,
let us note that the free energy density $\tilde{F}(D_\text{cut})$
at finite $D_\text{cut}$ and $V=1$ has the large-$N$ behavior
\begin{align}
  \label{F-Dcut1}
  \tilde{F}(D_\text{cut})&= C^{(0)}+\frac{1}{N^2} f(D_\text{cut})
  + \cdots \ , \\
    f(D_\text{cut})&\equiv
    \log\sum_{i=1}^{D_\text{cut}}  \exp(C_{r_i}^{(1)}) \ ,
\end{align}
as one can see from \eqref{eq:EK_volume_independent}.
In Fig.~\ref{fig:DcutDependence} (Right)
we plot
$N^2 \{ \tilde{F}(D_\text{cut}) - \tilde{F}(1) \}$ against $D_\text{cut}$,
where the results are seen to approach a single curve in the large-$N$
limit\footnote{This curve is not common to the U($N$) and SU($N$) gauge theories
  for $\lambda=0.5$ used here. However, it is common
  for $\lambda > 2$ due to the agreement of the coefficients $C_r^{(1)}$
  in the two theories.},
which corresponds to
the function $f(D_\text{cut})$ that appears in \eqref{F-Dcut1}.
 Note that the equivalence between
    the U($N$) and SU($N$) gauge theories at $V=1$ holds in the $D_{\rm cut} \rightarrow \infty$
    limit but not for finite $D_{\rm cut}$. In fact, the finite $D_{\rm cut}$ effects are smaller
    for SU($N$) with the same $N$, as one can see from Fig.~\ref{fig:DcutDependence} (Right),
    since the sum over the charge is taken already in \eqref{eq:character_coefficient}.




\subsection{Gross-Witten-Wadia phase transition}


As a demonstration of how the TRG works, let us discuss
the Gross-Witten-Wadia phase transition \eqref{eq:GWW_vacuum_energy},
which occurs in the large-$N$ limit.
%
By taking the derivative of the free energy density
\eqref{eq:EK_volume_independent},
we define the average plaquette and the specific heat as
\begin{align}
  W=-\frac{\lambda^2}{2}\frac{\partial F}{\partial\lambda} \ , \quad \quad
C
=-2\frac{\partial W}{\partial\lambda} \ ,
  \label{eq:GWW_plaquette}
\end{align}
whose large-$N$ limits are given by \cite{Gross1980,Wadia1980}
\begin{align}
     \lim_{N \rightarrow \infty} W  =\left\{
  \begin{array}{ll}
    \displaystyle 1-\frac{\lambda}{4}& (\lambda<2) \ , \\[2mm]
\displaystyle \frac{1}{\lambda}& (\lambda\geq2) \ ,
    \end{array}
  \right.
  \quad \quad
  \lim_{N \rightarrow \infty} C
  =\left\{
\begin{array}{ll}
      \displaystyle \frac{1}{2}&(\lambda<2) \ , \\[2mm]
    \displaystyle \frac{2}{\lambda^2}&(\lambda\geq2) \ .
    \end{array}
\right.
\label{eq:GWW_heat_capacity}
\end{align}



In order to calculate observables, we take the derivatives
of the partition function numerically using
the 9-point central finite difference schemes \cite{Fornberg1988}
\begin{alignat}{3}
    \epsilon f'(x) &=\tfrac{4}{5}f(x+\epsilon)
    -\tfrac{1}{5}f(x+2\epsilon)
    +\tfrac{4}{105}f(x+3\epsilon)
    -\tfrac{1}{280}f(x+4\epsilon)   \nonumber \\
    &\quad-\tfrac{4}{5}f(x-\epsilon)
    +\tfrac{1}{5}f(x-2\epsilon)
    -\tfrac{4}{105}f(x-3\epsilon)
    +\tfrac{1}{280}f(x-4\epsilon) \ ,\\
    \epsilon^2 f''(x) &=-\tfrac{205}{72}f(x)
    +\tfrac{8}{5}f(x+\epsilon)
    -\tfrac{1}{5}f(x+2\epsilon)
    +\tfrac{8}{315}f(x+3\epsilon)
    -\tfrac{1}{560}f(x+4\epsilon) \nonumber \\
    &\quad+\tfrac{8}{5}f(x-\epsilon)
    -\tfrac{1}{5}f(x-2\epsilon)
    +\tfrac{8}{315}f(x-3\epsilon)
    -\tfrac{1}{560}f(x-4\epsilon)
\end{alignat}
with $\epsilon=0.01$.

\mycomment{
  Specifically, we employ 9-point central finite difference schemes
  \cite{Fornberg1988} to evaluate first and second derivatives in our calculations:
\begin{equation}
    \begin{array}{rl}
    \epsilon f'(x)&\!\!\!=\tfrac{4}{5}f(x+\epsilon)
    -\tfrac{1}{5}f(x+2\epsilon)
    +\tfrac{4}{105}f(x+3\epsilon)
    -\tfrac{1}{280}f(x+4\epsilon)\\
    &\quad-\tfrac{4}{5}f(x-\epsilon)
    +\tfrac{1}{5}f(x-2\epsilon)
    -\tfrac{4}{105}f(x-3\epsilon)
    +\tfrac{1}{280}f(x-4\epsilon),\\
    \end{array}
\end{equation}
\begin{equation}
    \begin{array}{rl}
    \epsilon^2 f''(x)&\!\!\!=-\tfrac{205}{72}f(x)
    +\tfrac{8}{5}f(x+\epsilon)
    -\tfrac{1}{5}f(x+2\epsilon)
    +\tfrac{8}{315}f(x+3\epsilon)
    -\tfrac{1}{560}f(x+4\epsilon)\\
    &\quad+\tfrac{8}{5}f(x-\epsilon)
    -\tfrac{1}{5}f(x-2\epsilon)
    +\tfrac{8}{315}f(x-3\epsilon)
    -\tfrac{1}{560}f(x-4\epsilon)
    \end{array}
\end{equation}
with $\epsilon=0.01$.
}

\begin{figure}
\centering
{\includegraphics[scale=0.8]{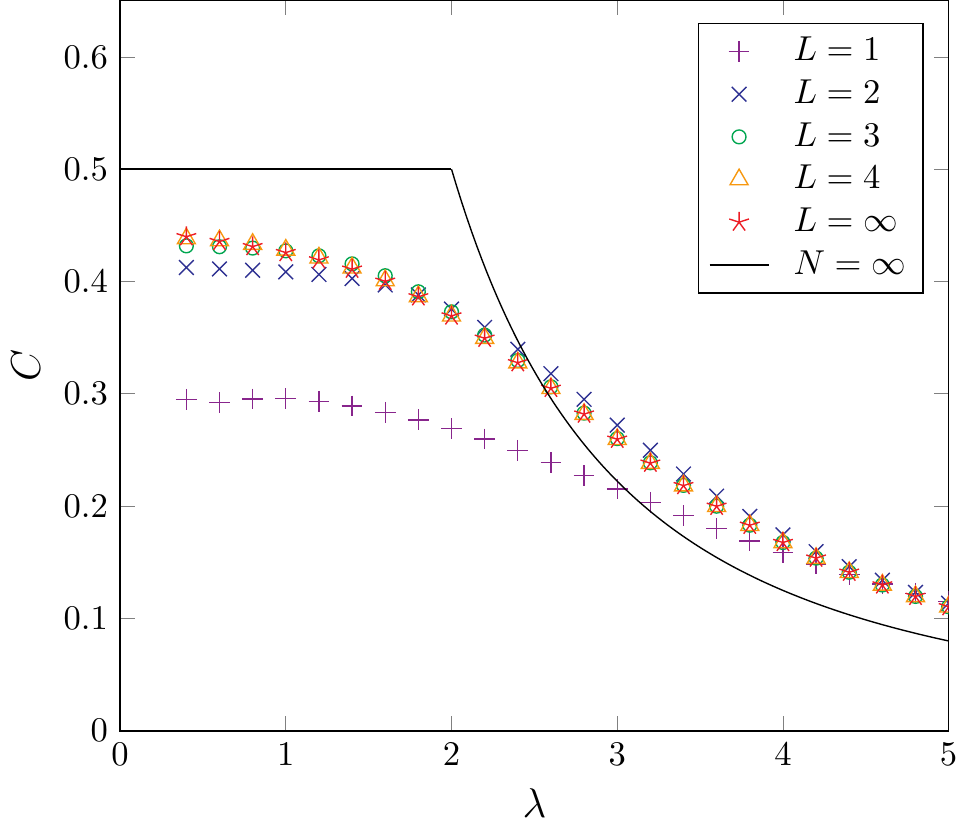}}
{\includegraphics[scale=0.8]{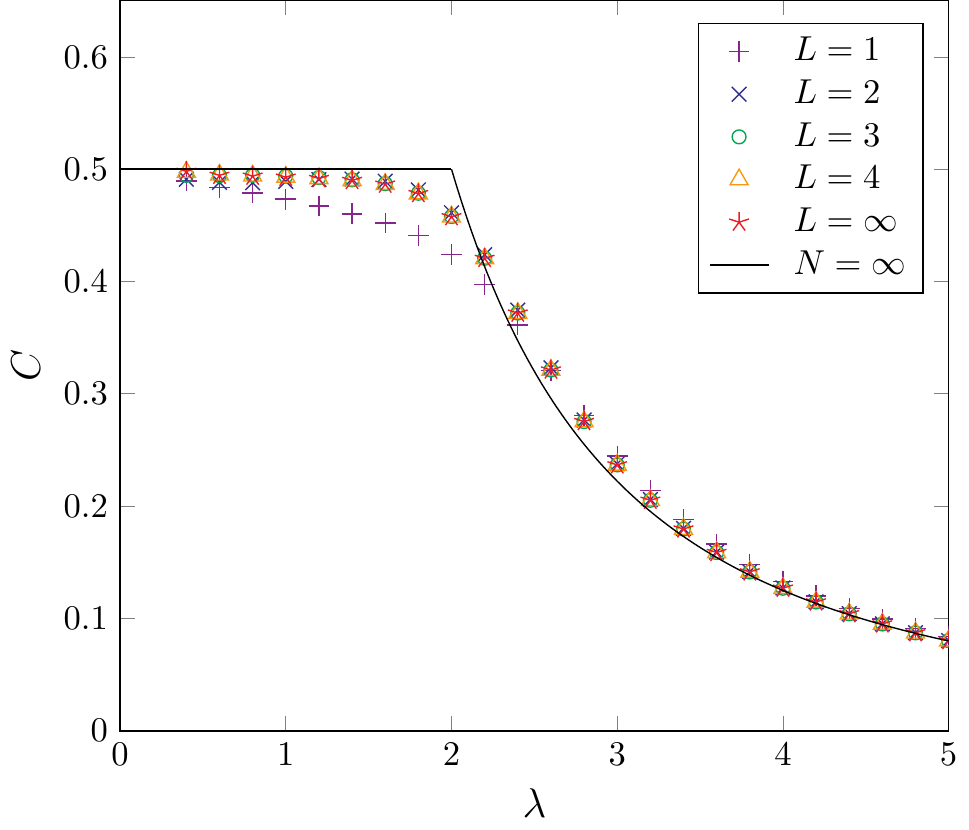}}
\caption{The specific heat $C$
  defined in \eqref{eq:GWW_plaquette}
  is plotted against the 't Hooft coupling constant $\lambda$
  for the $\G{SU}(3)$ (Left) and $\G{SU}(10)$ (Right) gauge theories with various $L$
  using $D_\text{cut}=64$.
  The solid line represents the large-$N$ limit obtained analytically
  as \eqref{eq:GWW_heat_capacity}.}
\label{fig:GWW}
\end{figure}

In Fig.~\ref{fig:GWW}, we plot
the specific heat obtained by differentiating
the free energy density
numerically
for the $\G{SU}(3)$ (Left) and $\G{SU}(10)$ (Right)
gauge theories
with various $L$.
The bond dimension is chosen to be $D_\text{cut}=64$.
We see that the finite volume effects are reduced considerably
by increasing $N$ from 3 to 10 due to the Eguchi-Kawai reduction.
We also see how the Gross-Witten-Wadia phase transition 
appears as $N$ increases.


\section{2D U($N$) gauge theory
  with the \texorpdfstring{$\theta$}{theta} term}
\label{section:ThetaTerm}

In this section, we show how the TRG works when we add a $\theta$ term
to the 2D gauge theory. Exact solutions are obtained and
discussed in Refs.~\cite{Bonati2019,Bonati2019a}.
This model is of particular interest since ordinary Monte Carlo simulation
becomes extremely difficult due to the sign problem.
Recently the complex Langevin method \cite{Hirasawa2020}
and the density-of-state method \cite{Gattringer:2020mbf} were
applied to this model successfully. (See also Ref.~\cite{Pawlowski:2021bbu}.)
In these studies, however, one has to dismiss periodic boundary conditions
in order to avoid the topology freezing problem, which refers to
the problem that one cannot sample configurations
in
different
topological sectors with the correct weight.
The TRG is not only free from the sign problem but also
free from the topology freezing problem.
It therefore allows us to obtain results even with periodic boundary conditions
unlike in Refs.~\cite{Hirasawa2020,Gattringer:2020mbf}.
In the $\G{U}(1)$ case, the TRG was applied to this model 
using the Gauss quadrature instead of the character expansion
in order to discretize the indices of the fundamental tensor \cite{Kuramashi:2019cgs}.


\subsection{the first-order phase transition at $\theta=\pm \pi$}

In two dimensions, the $\theta$ term in the continuum action reads
\begin{align}
  \label{theta-action}
  S_\theta&=-i\, \theta \, Q \ ,
  \\
        Q&=\frac{1}{4\pi}\int d^2x \, \epsilon_{\mu\nu} \, \trace F_{\mu\nu} \ ,
    \label{top-charge-cont}
\end{align}
where the topological charge $Q$ takes integer values on a compact manifold.
Since the topological charge $Q$ vanishes identically
for the $\G{SU}(N)$ case in 2D,
we restrict ourselves to the $\G{U}(N)$ case in this section.
Putting the theory on a periodic lattice,
the action is given by the plaquette action \eqref{plaquette-action}
with an additional term \eqref{theta-action},
where the definition of the topological charge $Q$
is given on the lattice by 
\begin{equation}
  Q=\frac{1}{2\pi i}\sum_n \log \det P_n \ .
  \label{eq:log_definition} 
\end{equation}
The complex log in \eqref{eq:log_definition}
is defined by using the principal value.
The topological charge $Q$ thus defined
takes integer values even before taking the continuum limit,
which guarantees that the
partition function
is invariant under $\theta \mapsto \theta + 2\pi$.

The exact solution for the theory with the $\theta$ term
can be obtained as in the $\theta=0$ case discussed in Section \ref{2dUNplusTheta}.
Corresponding to \eqref{eq:analytic_partition_function}, one obtains \cite{Bonati2019}
\begin{align}
    \label{eq:paritition_function_theta}
  Z(\theta)& =
  \sum_r
   \left( \frac{\tilde\gamma_r(\theta)}{d_r} \right)
   ^{L_1 L_2} \ , \quad \quad 
  \\
  \tilde\gamma_r(\theta)&=\det\mathcal{M}_{r}(\theta) \ ,
  \label{eq:character_coefficient_theta}\\
                [\mathcal{M}_{r}(\theta)]_{ij}&=
                \int_{-\pi}^{+\pi}\frac{d\phi}{2\pi}\cos \{ (l_j+i-j+\tfrac{\theta}{2\pi})\phi \}
                \exp \left(\tfrac{2N}{\lambda}\cos\phi \right) \ ,
    \label{eq:character_coefficient_determinant_theta}
\end{align}
%
and corresponding to \eqref{def-Gi},
the singular values become
\begin{equation}
\sigma_r (\theta) = \frac{|\tilde\gamma_r(\theta)|}{d_r} \   .
  \label{eq:fundamental_tensor_theta}
\end{equation}
Note that \eqref{eq:character_coefficient_determinant_theta}
implies that the singular values have the properties
\begin{align}
    \label{eq:sigma-relation-theta1}
  \sigma_{(r^{\rm (U)}, \, q)}(\theta+2\pi) &= \sigma_{(r^{\rm (U)},\, q+1)}(\theta) \ , \\
  \sigma_{r^{\rm (U)}}(-\theta) &= \sigma_{\bar{r}^{\rm (U)}}(\theta) \ .
    \label{eq:sigma-relation-theta2}
\end{align}
Namely the singular values are not invariant under
$\theta \mapsto \theta + 2 \pi$ or $\theta \mapsto - \theta$,
while
the partition function is, because of the summation over
representations $r$ involved in \eqref{eq:paritition_function_theta}.

\begin{figure}
\centering
\includegraphics[scale=0.7]{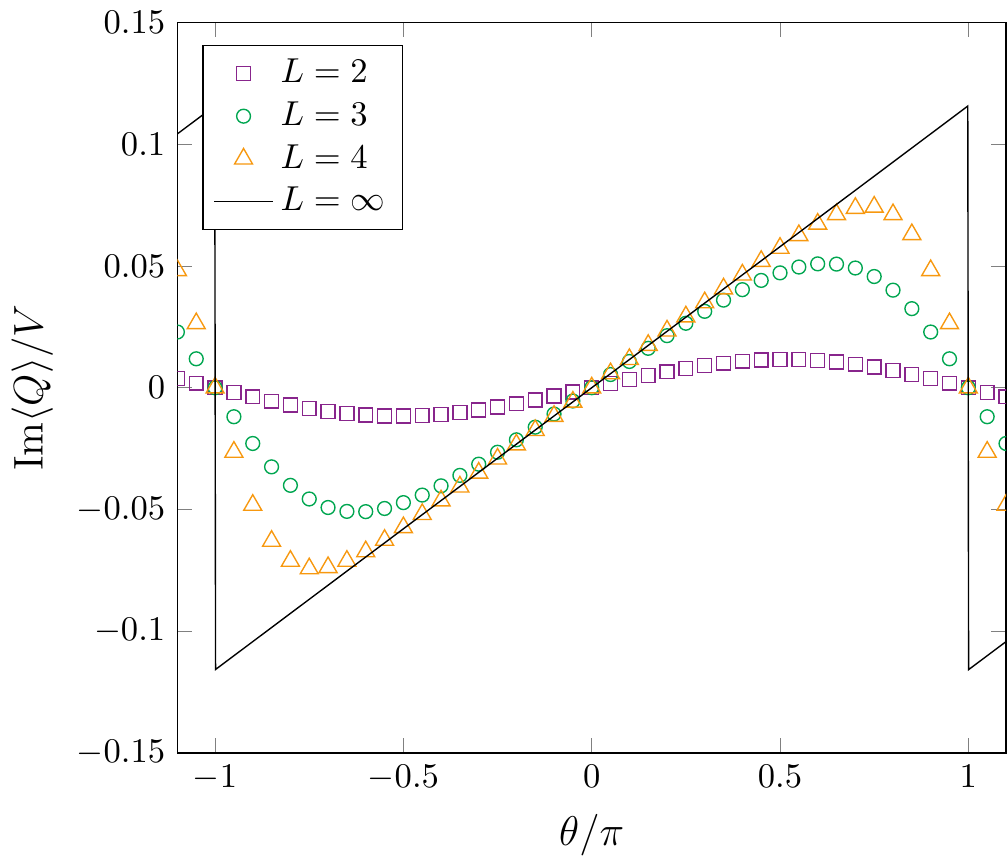}
\includegraphics[scale=0.7]{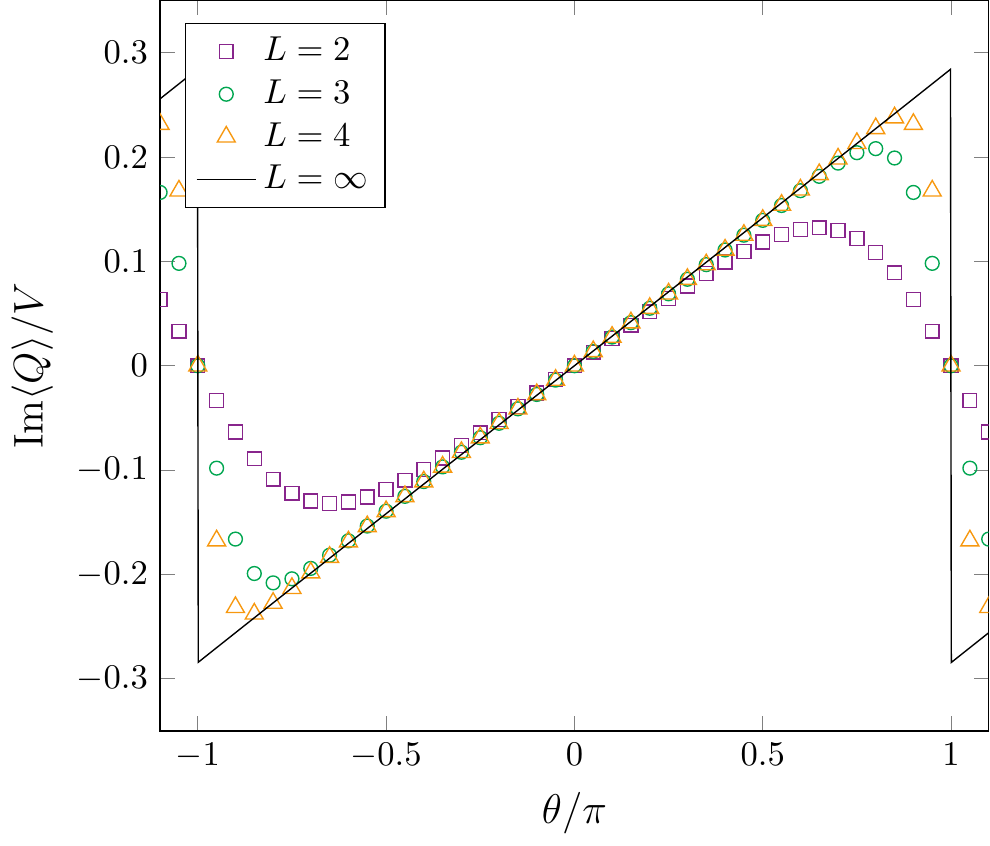}
\caption{The imaginary part of the topological charge density is
  plotted against $\theta$
  for the U($3$) gauge theory with various $L$, $D_\text{cut}=32$
  using $\lambda=1.5$ (Left) and $\lambda=3$ (Right).
  A clear gap is seen to develop at $\theta=\pm\pi$ with increasing $L$.}
\label{fig:ImQ}
\end{figure}

The most interesting feature of this theory
is the existence of the first-order phase transition at $\theta=\pi$,
which is associated with the spontaneous breaking of the parity symmetry.
In order to see this, let us first note
that the expectation value of the topological charge $Q$
is pure imaginary for arbitrary $\theta$
due to the fact that $Q$ is real and parity odd.
We therefore define the topological charge density as
\begin{align}
  \frac{1}{V} \text{Im} \langle  Q\rangle
  &= -\frac{1}{V} \frac{\partial}{\partial \theta} \log Z(\theta) \ ,
    \label{eq:ImQ-partition-fn}
\end{align}
which develops a gap
at $\theta=\pm(\pi-\epsilon)$
for $\epsilon \rightarrow +0$
as the volume $V$ increases.
%
In Fig.~\ref{fig:ImQ} we plot
the topological charge density\footnote{We find
  that the topological charge vanishes identically for $L=1$.
  This is expected from the fact that
the one-site model
cannot distinguish U($N$) and SU($N$)
as we explained
in Section \ref{subsection:EK-equiv}.} against $\theta$
for the U($3$) gauge theory with various $L$
using $D_\text{cut}=32$ and $\lambda=3$,
which exhibits a clear gap at $\theta=\pm \pi$
with increasing $L$.

From now on, we discuss the large-$N$ behavior of the theory
with the $\theta$ term.
Let us first mention that the topological susceptibility defined by
\begin{align}
  \chi
&=   - \lim_{V\rightarrow \infty} \frac{1}{V} \left.
  \frac{\partial^2}{\partial \theta ^2} \log Z(\theta) \right|_{\theta=0}
  =
  \lim_{V\rightarrow \infty}
  \left.\frac{1}{V}
  \Big(\langle Q^2\rangle-\langle Q\rangle^2\Big)\right|_{\theta=0}
  \label{eq:top-susc-paritition_function_theta}
\end{align}
is obtained at large $N$ as \cite{Bonati2019}
\begin{align}
 \chi  &\sim \chi_N \equiv  \left\{
    \begin{array}{ll}
    \displaystyle -\frac{1}{4\pi^2}\log\left(1-\frac{\lambda}{2}\right)& \mbox{~for~}\lambda<2 \ ,\\[5mm]
    \displaystyle \frac{1}{2\pi^2} \left\{ \log N+
      \log \left(1-\frac{2}{\lambda} \right)+\gamma_\text{E}+1 \right\} & \mbox{~for~}\lambda>2 \ ,
    \end{array}
    \right.
    \label{eq:susceptibility_predictions}
\end{align}
where $\gamma_\text{E}\approx0.5772$ is the Euler constant.
Notice the appearance of $\log N$ for $\lambda >2$.
Since $\text{Im} \langle Q\rangle/V \sim \chi_N \theta$ at small $\theta$,
we consider the quantity $\text{Im}\langle Q\rangle/(V \chi_N)$,
which is normalized by $\chi_N$, in order to make the quantity finite
in the $N\rightarrow \infty$ limit.

\begin{figure}
\centering
\includegraphics[scale=0.8]{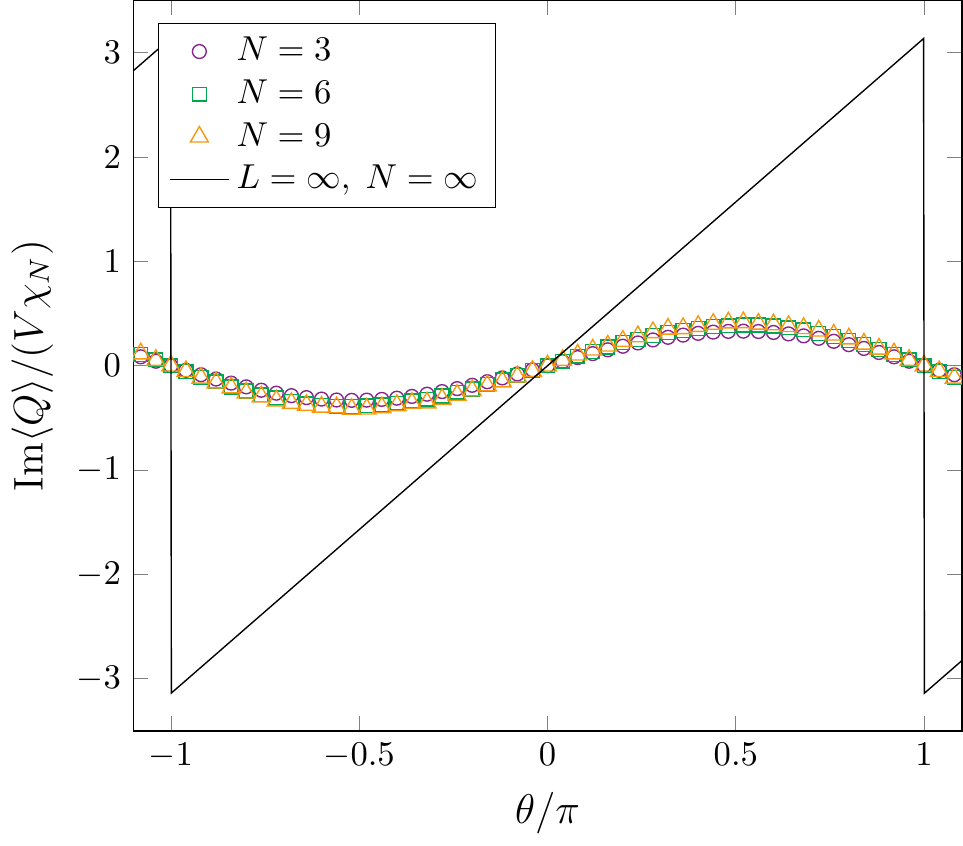}
\includegraphics[scale=0.8]{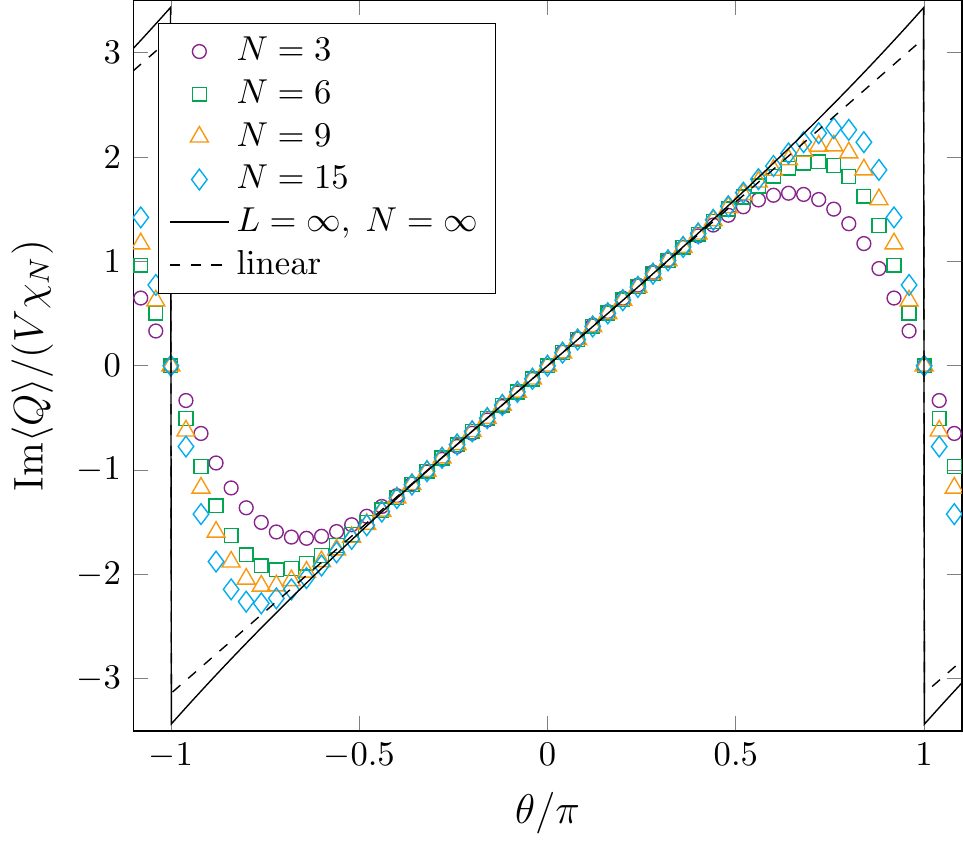}
\caption{The imaginary part of the topological charge density is plotted
  against $\theta$
  for U($N$) gauge theories with various $N$ and
  $L=2$, $D_\text{cut}=32$ using $\lambda=1.5$ (Left) and $\lambda=3$ (Right).
  The results for $N=\infty$ and $L=\infty$ are shown by the solid lines,
  which, for $\lambda=3$, deviates from the linear behavior 
  represented by the dashed line in the right panel.
}
\label{fig:ImQLargeN}
\end{figure}

In Fig.~\ref{fig:ImQLargeN}, we plot
$\text{Im}\langle Q\rangle/(V \chi_N)$ against $\theta$
in the U($N$) gauge theory for various $N$
with $L=2$, $D_\text{cut}=32$ using $\lambda=1.5$ (Left) and
$\lambda=3$ (Right).
We find that the results for $\lambda=3$ approach the results
for $L=\infty$ as $N$ increases, whereas the results for $\lambda=1.5$
do not\footnote{We also find in Fig.~\ref{fig:ImQLargeN}
  that the results for $\lambda=1.5$ 
  become completely linear in $\theta$ 
  in the $L \rightarrow \infty$ and $N \rightarrow \infty$ limits,
  whereas the results for $\lambda=3$ do not.
  This can be understood from the
  large-$N$ behavior of the singular values given by
  \eqref{eq:sigma-theta-largeN-WC} and
  \eqref{eq:sigma-theta-largeN-SC} for
  $\lambda < 2$ and $\lambda > 2$, respectively.}.
This suggests that the volume independence holds
in the large-$N$ limit in the strong coupling phase.
In fact,
  we see in Fig.~\ref{fig:ImQ} that
  finite volume effects are considerably reduced for $\lambda=3$
  compared with $\lambda=1.5$, which
  shows some tendencies towards
  the volume independence for $\lambda>2$ already at $N=3$.
  We will provide a theoretical understanding of this
  novel volume independence
  based on the large-$N$ behavior of the singular-value spectrum
  in the presence of the $\theta$ term.

\subsection{the large-$N$ behavior of the singular-value spectrum}
\label{sec:SV-largeN-theta}

Let us recall here that
the summation over representations in \eqref{eq:paritition_function_theta}
is dominated by the trivial representation
in the $V \rightarrow \infty$ limit for any $\theta$ in $|\theta|<\pi$.
Therefore, \eqref{eq:susceptibility_predictions}
implies
\begin{align}
  \log  \sigma_{{\rm trv}^{\rm (U)}}(\theta) 
& \sim
\log  \sigma_{{\rm trv}^{\rm (U)}}(0)  - \frac{1}{2} \chi_N \theta^2 
    \label{eq:top-susc-paritition_function_theta2}
\end{align}
at $\theta \sim 0$.
Note that
  $\sigma_r(\theta)$ is an even function of $\theta$
  for real representations $r$ in general due to the property
  \eqref{eq:sigma-relation-theta2}.
  In what follows, we investigate
the large-$N$ properties of the singular values for
general representations.

\begin{figure}
\centering
{\includegraphics[scale=0.7]{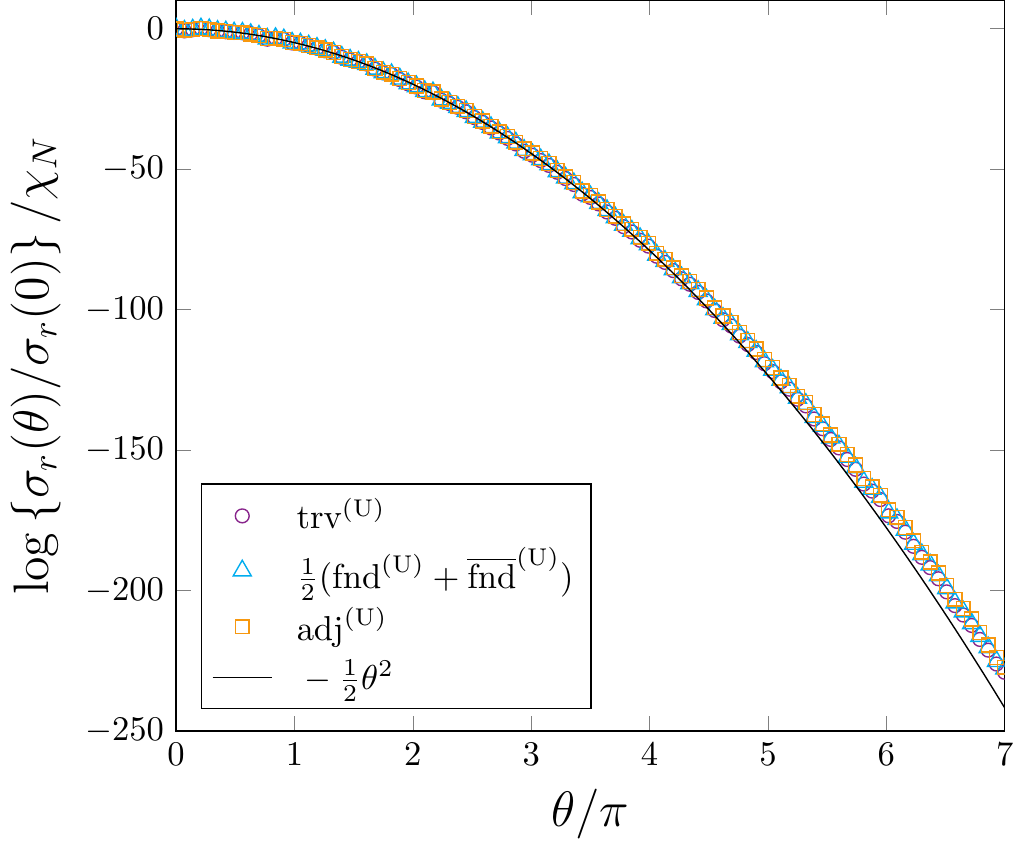}}
{\includegraphics[scale=0.7]{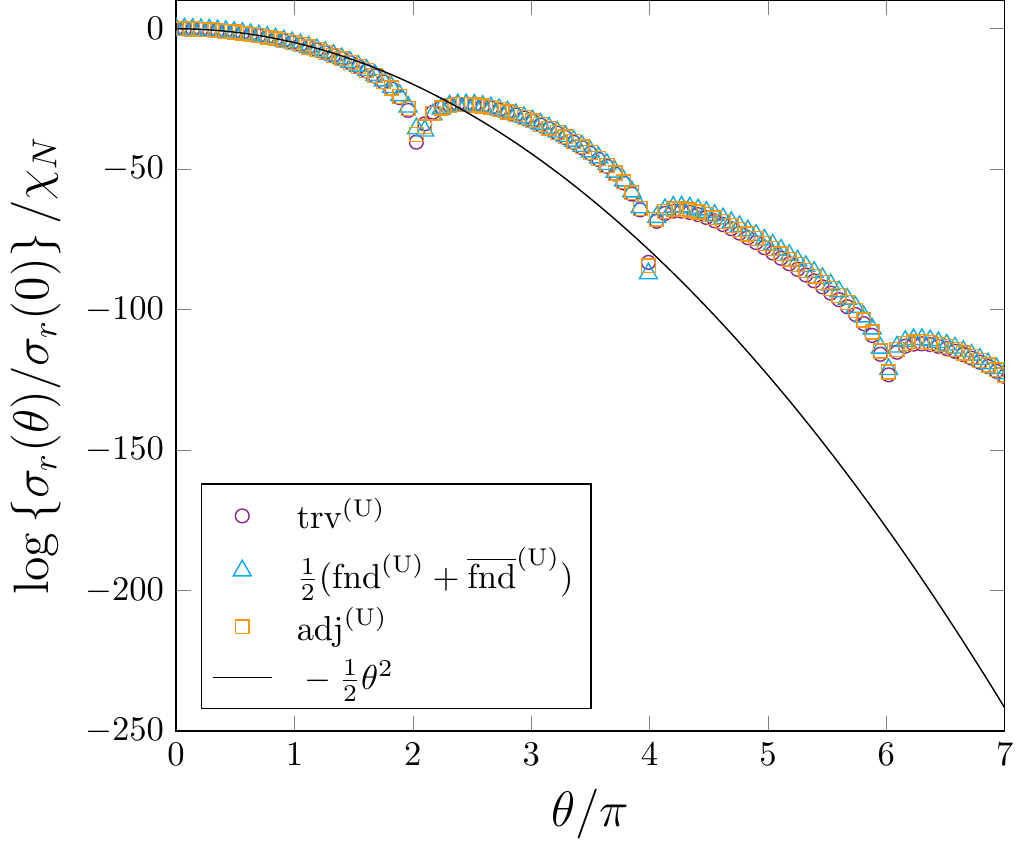}}
{\includegraphics[scale=0.7]{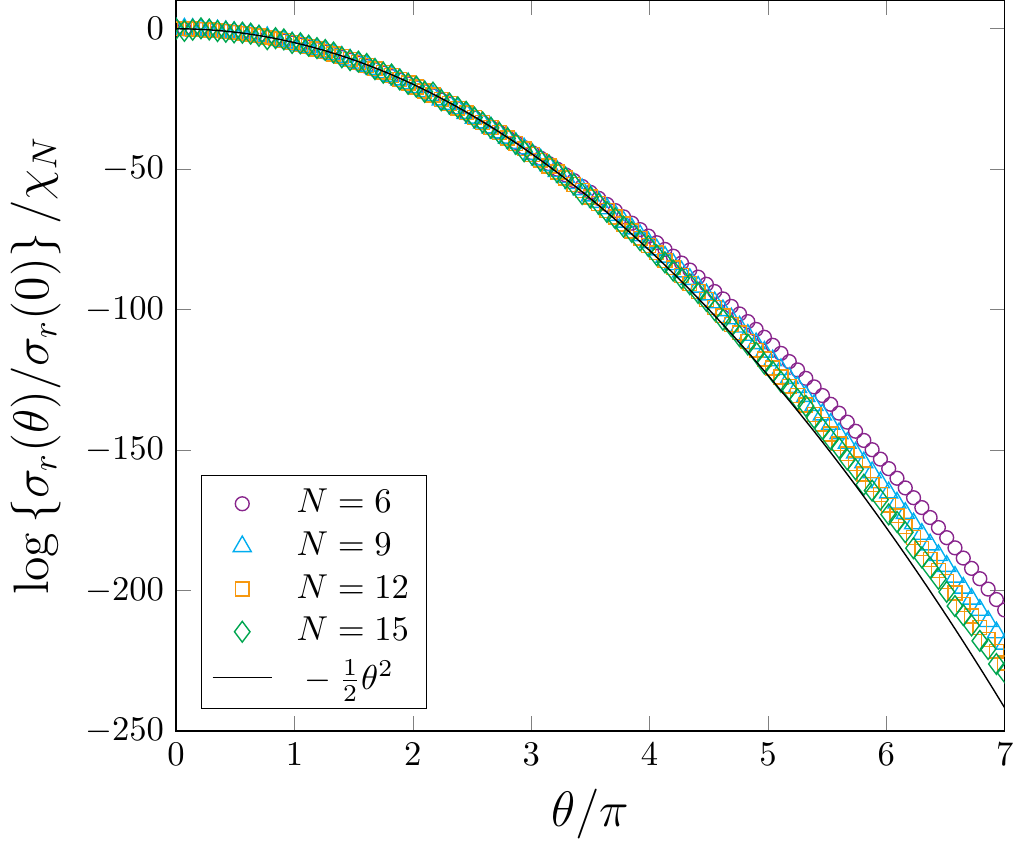}}
{\includegraphics[scale=0.7]{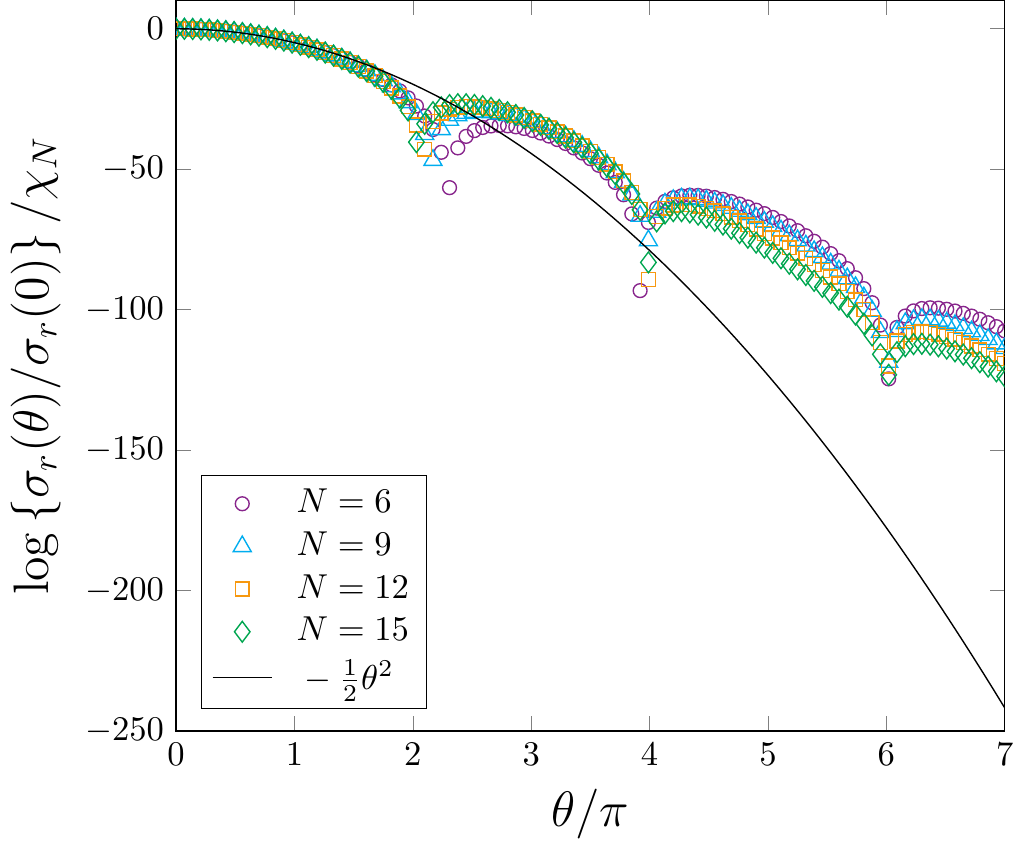}}
\caption{(Top) The quantity
  $\frac{1}{\chi_N}\log \{ \sigma_r(\theta)/\sigma_r(0) \}$
  is plotted against
    $\theta$ for
  the trivial, fundamental and adjoint representations
  in the U($N$) gauge theory with $N=15$ at $\lambda=1.5$ (Left)
  and $\lambda=3$ (Right).
  We also show the behavior $-\frac{1}{2} \theta^2$ by the
  solid lines for comparison.
  (Bottom)
  The same plots with various $N$ at $\lambda=1.5$ (Left)
  and $\lambda=3$ (Right)
  are shown for the trivial representation.
  A clear large-$N$ scaling behavior is seen.
  Similar behaviors are
  observed for the other representations.
}
\label{fig:universality}
\end{figure}

\begin{figure}
\centering
\includegraphics[scale=0.8]{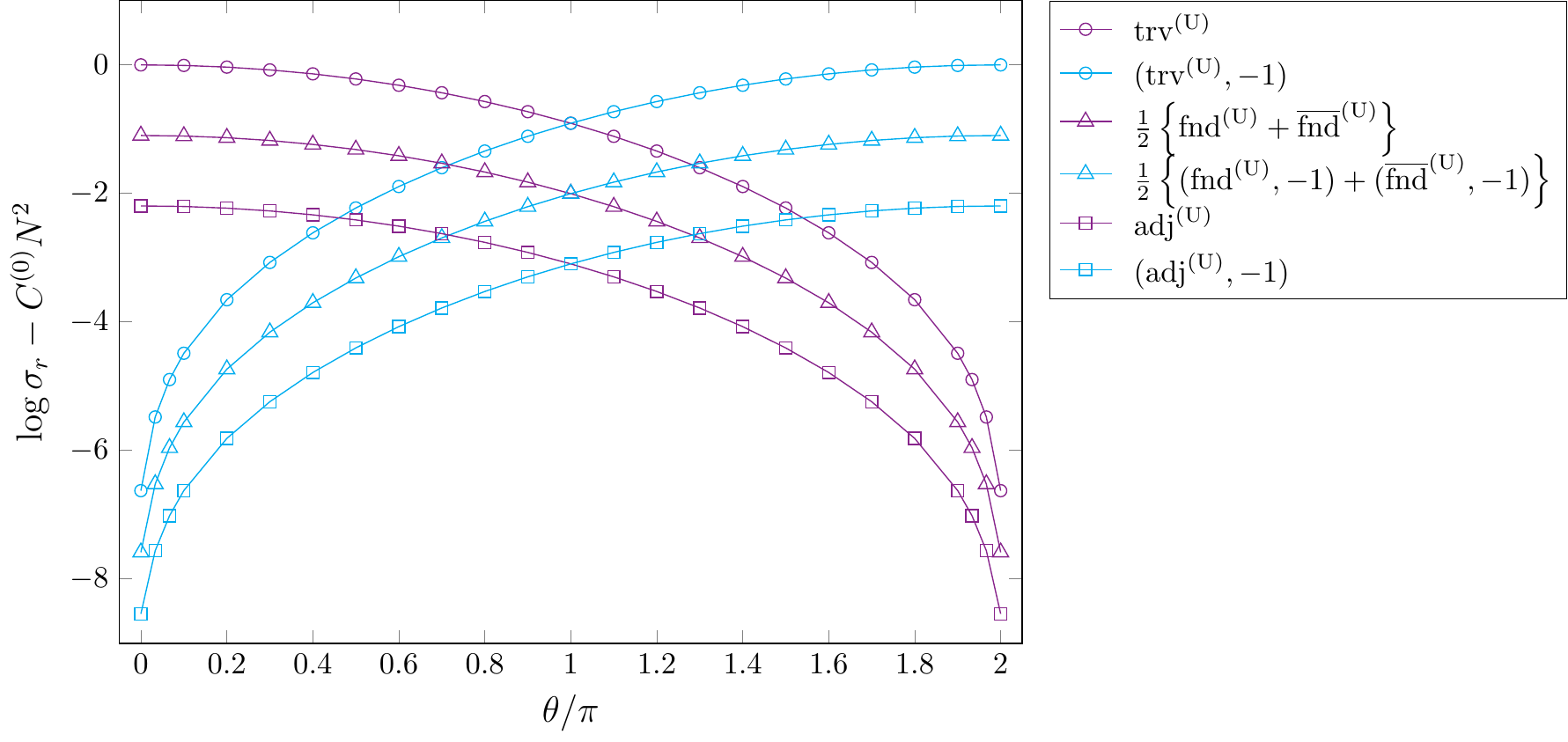}
\caption{The quantity
  $(\log(\sigma_r)-C^{(0)} N^2)$
  is plotted against $\theta$ for
  trivial, fundamental and adjoint representations
  in the U($N$) gauge theory with $N=20$ at $\lambda=3$.
  We also plot the results for representations with charge $q=-1$
  with blue symbols.}
\label{fig:SVvTheta}
\end{figure}

Let us first discuss
the weak coupling phase $\lambda < 2$.
For the sequence \eqref{sequence-UN} of representations
without charge,
we find that\footnote{A linear term in $\theta$ appears
  at O($1/N$) for complex representations
  but not for the real ones.}
\begin{align}
  \log \sigma_{r^{\rm (U)}}(\theta)
  &=
  \log \sigma_{r^{\rm (U)}}(0)
  - \frac{1}{2} \chi_N \theta^2 + {\rm O}\left(\frac{1}{N}\right)
    \label{eq:sigma-theta-largeN-WC}
\end{align}
at large $N$,
where
$\chi_N$ is an ${\rm O}(1)$ quantity given
by \eqref{eq:susceptibility_predictions}.
Note that the singular values $\sigma_r(\theta)$
are not periodic in $\theta$ as we mentioned
below \eqref{eq:sigma-relation-theta1},
and \eqref{eq:sigma-theta-largeN-WC} holds for arbitrary $\theta$.
This is shown in Fig.~\ref{fig:universality} (Top-Left)
for the trivial,
fundamental\footnote{Since the fundamental representation is not
  real unlike the trivial and adjoint representations,
  the $1/N$ terms in \eqref{eq:sigma-theta-largeN-WC}
  involves a linear term in $\theta$.
  Here we take an average of the results for
  the fundamental and anti-fundamental representations
  to cancel this term for the sake of simplicity.
  This remark also applies to
  Fig.~\ref{fig:SVvTheta}, where we add 
  charge $-1$ to the fundamental representation.} and
adjoint representations.
The small deviation at large $\theta$ is 
due to finite $N$ effects as we
can see from Fig.~\ref{fig:universality} (Bottom-Left).

When we add some charge $q$ to the representation
$r^{\rm (U)}$, we obtain
\begin{align}
   \sigma_{(r^{\rm (U)}, \, q)}(\theta) &=
   \sigma_{r^{\rm (U)}}(\theta+2\pi q)
   = \sigma_{r^{\rm (U)}}(0) e^{-\frac{1}{2} \chi_N (\theta+2\pi q)^2 + {\rm O}(1/N)} 
    \label{eq:sigma-theta-largeN2}
\end{align}
using \eqref{eq:sigma-relation-theta1} and
\eqref{eq:sigma-theta-largeN-WC}.
Setting $\theta=0$ in \eqref{eq:sigma-theta-largeN2}, we get
\begin{align}
   \sigma_{(r^{\rm (U)}, \, q)}(0) &=
   \sigma_{r^{\rm (U)}}(0) e^{- 2 \pi^2  q^2 \chi_N + {\rm O}(1/N)}  \ ,
    \label{eq:sigma-theta-largeN3}
\end{align}
which provides a clear understanding of the
results in Fig.~\ref{fig:SVvsN2UN} (Top),
where
$\log \sigma_r$
for representations with some charge $q$ has a slope
commonly decreased by $2 \pi ^2 q^2 \chi_N $.
When we switch on $\theta$,
\eqref{eq:sigma-theta-largeN2}
implies that
the singular-value spectrum still has a definite profile
in the large-$N$ limit,
and the $\theta$-dependence of the profile comes only from
the factor $e^{-\frac{1}{2} \chi_N (\theta+2\pi q)^2}$,
which depends on
the charge $q$.

Let us next discuss
the strong coupling phase $\lambda > 2$, where
$\chi_N$ involves a $\log N$ term as in
\eqref{eq:susceptibility_predictions}.
For the sequence \eqref{sequence-UN} of representations without charge,
we observe from Fig.~\ref{fig:universality} (Top-Right) that
\begin{align}
  \log \sigma_{r^{\rm (U)}}(\theta)
  &=
  \log \sigma_{r^{\rm (U)}}(0)
  - \chi_N \varphi(\theta) + {\rm O}\left(\frac{1}{N}\right)
    \label{eq:sigma-theta-largeN-SC}
\end{align}
 at large $N$ except for $\theta \sim 2 n \pi$
($|n|=1, 2, \cdots $).
%
%
The function $\varphi(\theta)$ in \eqref{eq:sigma-theta-largeN-SC}
is
given by some even function of $\theta$, which is independent of $r^{\rm (U)}$
and satisfy
$\varphi(\theta) \sim  \frac{1}{2} \theta^2$ at small $\theta$.
%

\begin{figure}
\centering
\includegraphics[scale=0.65]{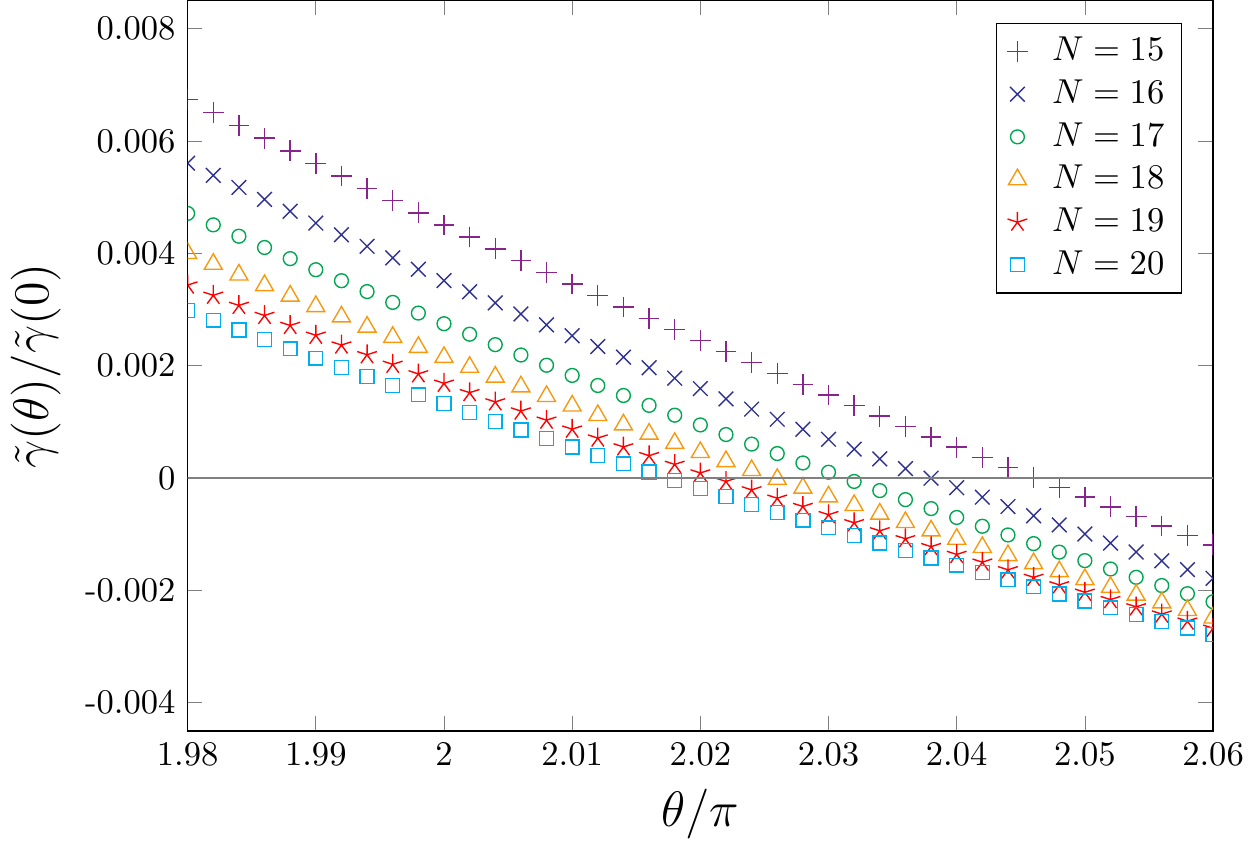}
\includegraphics[scale=0.65]{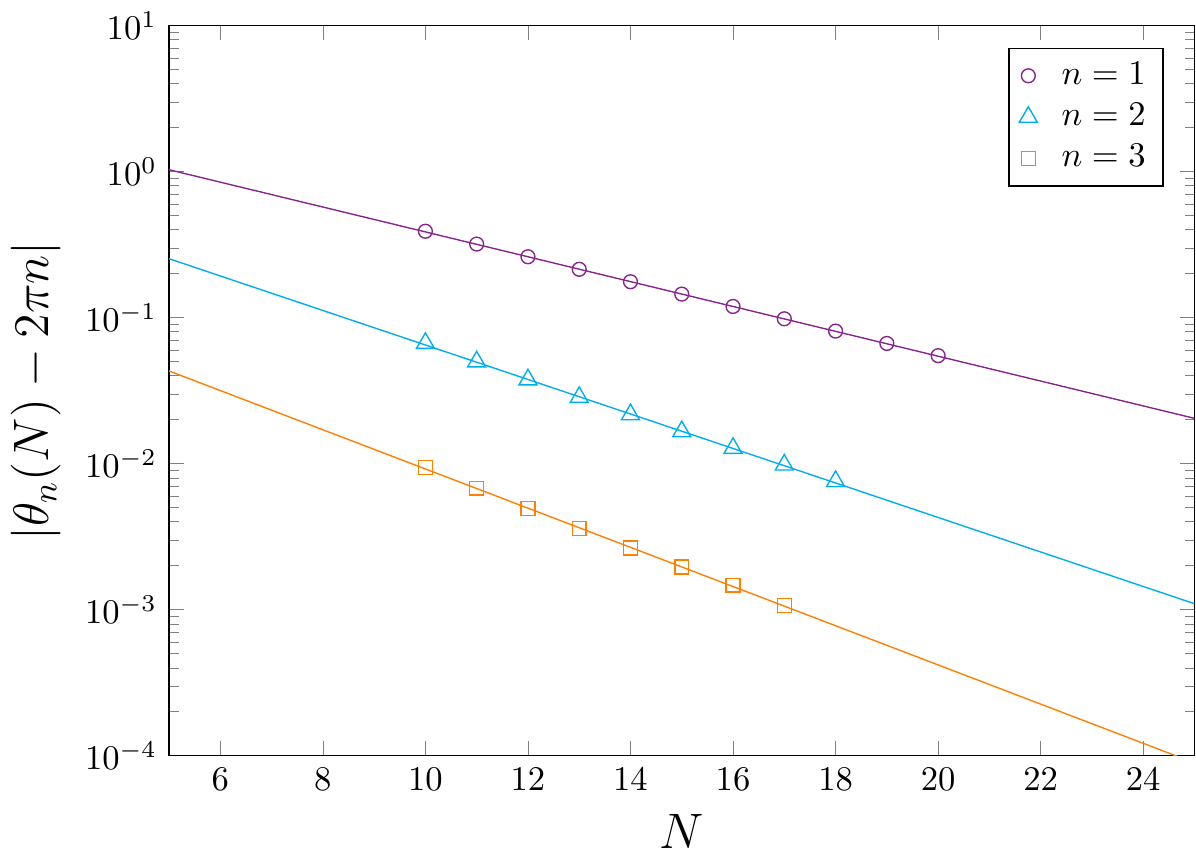}
\includegraphics[scale=0.65]{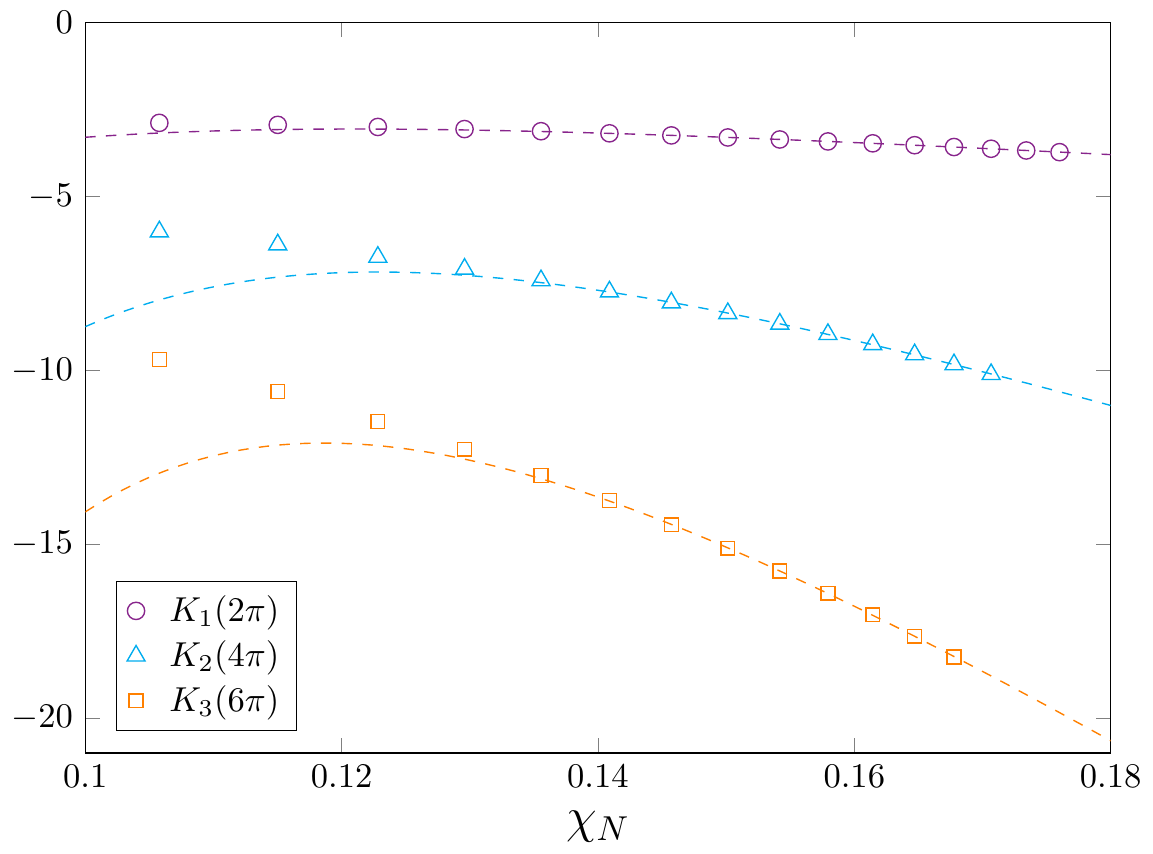}
\caption{(Top) The ratio
  $\tilde{\gamma}_{\rm trv} (\theta) / \tilde{\gamma}_{\rm trv}(0)$
  for the trivial representation is plotted against
    $\theta$
  in the U($N$) gauge theory with various $N$ at $\lambda=3$.
  (Bottom-Left) The deviation of the zeroes $\theta_n(N)$
  from $2\pi n$ is plotted against $N$ for $n=1,2$ and 3 in the log scale.
  The solid lines are fits to the exponential
  behavior $|\theta_n(N)-2\pi n| \sim A_n e^{- B_n N}$.
  (Bottom-Right)
  The quantity $K_n(2\pi n)$
  is plotted for $n=1,2$ and 3
  against $\chi_N$.
  The dashed lines are fits to the behavior
  $K_n(2 \pi n) = b_n \chi_N + c_n + d_n /N^2 $.
}
\label{fig:zeroes_scaling}
\end{figure}

At $\theta \sim 2 n \pi$ ($|n|=1, 2, \cdots $),
the function $\varphi(\theta)$ has singularities,
which appear because
$\tilde{\gamma}_{r^{\rm (U)}}(\theta)$
in \eqref{eq:fundamental_tensor_theta}
oscillates around zero as a function of $\theta$.
In the vicinity of these zeroes, which we denote by $\theta_n(N)$, we find that
\begin{align}
\sigma_{r^{\rm (U)}}(\theta)
 &\sim  \sigma_{r^{\rm (U)}}(0) \, 
 e^{b_n \chi_N+ v_n } |\theta - \theta_n (N)| \ ,
 \label{eq:sigma-theta-largeN-singularity} \\
 |\theta_n(N) - 2\pi n| & \sim e^{a_n N+u_n} \quad \quad  (a_n < 0) 
 \label{eq:sigma-theta-largeN-singularity-zeroes} 
\end{align}
at large $N$.
In Fig.~\ref{fig:zeroes_scaling}, we confirm this behavior
at $\lambda=3$
for the trivial representation, but the same scaling behavior
is confirmed also for the fundamental and adjoint representations.
In the top panel, we plot
$\tilde{\gamma}_{\rm trv}(\theta)/\tilde{\gamma}_{\rm trv}(0)$
against
$\theta$,
which shows that
the ratio
crosses zero at $\theta=\theta_n(N)\sim 2\pi n$ for $n=1$.
In the bottom-left panel,
we plot $|\theta_n(N) - 2\pi n|$ against $N$ in the log scale,
which confirms \eqref{eq:sigma-theta-largeN-singularity-zeroes}
for $n=1,2$ and 3.
In the bottom-right panel, we define the function $K_n(\theta)$ by
\begin{align}
\sigma_{r^{\rm (U)}}(\theta)
 &=  \sigma_{r^{\rm (U)}}(0) \, 
 e^{K_n(\theta)} |\theta - \theta_n (N)| 
 \label{eq:sigma-theta-largeN-singularity-K}
\end{align}
around $\theta \sim \theta_n(N)$
and plot $K_n(2\pi n)$ 
against $\chi_N$ for $n=1,2$ and 3,
which confirms
\eqref{eq:sigma-theta-largeN-singularity}
at large $N$.

Adding some charge $q\neq 0$ to the representation
$r^{\rm (U)}$ and setting $\theta=0$, we obtain 
\begin{align}
  \sigma_{(r^{\rm (U)}, \, q)}(0)
  &=    \sigma_{r^{\rm (U)}}(2 \pi q)
  = \sigma_{r^{\rm (U)}}(0) \, e^{a_q N + b_q \chi_{N} + c_q } \ ,
    \label{eq:sigma-theta-largeN2-SC}
\end{align}
using \eqref{eq:sigma-relation-theta1},
\eqref{eq:sigma-theta-largeN-singularity}
and \eqref{eq:sigma-theta-largeN-singularity-zeroes},
where we have defined $c_n=u_n+v_n$.
This implies
\begin{align}
\log \sigma_{(r^{\rm (U)}, \, q)}(0)
  &=  \log \sigma_{r^{\rm (U)}}(0) + a_q N + b_q \chi_N + c_q  \ ,
    \label{eq:sigma-theta-largeN2-SC-abc}
\end{align}
which provides a clear understanding of
our results in Fig.~\ref{fig:SV-subtract}.
As we discussed in Section \ref{subsection:dependence-coupling},
the representations with some charge have a negative O($N$) term
with the same coefficient for the same charge and they all
disappear from the spectrum at large $N$.

Let us discuss what happens for $\theta \neq 0$.
Using
\eqref{eq:sigma-relation-theta1}
and \eqref{eq:sigma-theta-largeN-SC},
we obtain 
\begin{align}
 \sigma_{(r^{\rm (U)}, \, q)}(\theta)
  &=
   \sigma_{r^{\rm (U)}}(0) \, 
  e^{- \chi_N \varphi(\theta+2\pi q) + {\rm O}(1/N) } \ .
    \label{eq:sigma-theta-largeN-SC-q}
\end{align}
This is confirmed in Fig.~\ref{fig:SVvTheta}, where we plot
$(\log \sigma_r-C^{(0)} N^2)$
against $\theta$ for various representations
in the U($N$) gauge theory with $N=20$ at $\lambda=3$.
We can see that the curves
have
the same $\theta$ dependence
for the same charge $q$
neglecting the ${\rm O}(1/N)$ terms in \eqref{eq:sigma-theta-largeN-SC-q}.
Note also that the representations with charge $q=-1$
take over at $\theta > \pi$,
which ensures the $2\pi$ periodicity of the partition function.

Eq.~\eqref{eq:sigma-theta-largeN-SC-q}
implies that
it is also true for nonzero $|\theta| < \pi$
that only
the sequences \eqref{sequence-UN} of representations
without charge
dominate in the large-$N$ limit at $\lambda >2$
since the coefficient of the negative O($\log N$) term in $\log \sigma_r$
is
larger in magnitude for $q\neq 0$. 
Moreover, since the $\theta$-dependence comes only from
the common overall factor $e^{-\chi_N \varphi(\theta)}$ for
the $q=0$ sector
that dominates in the large-$N$ limit,
the singular-value spectrum has a definite profile in the large-$N$ limit,
which is $\theta$-independent.
Let us recall that the profile
at $\theta=0$
is the same as that for SU($N$) in the $\lambda>2$ region
as we discussed in Section \ref{subsection:dependence-coupling}.
This profile remains unaltered at nonzero $\theta$
up to the $\theta$-dependent overall factor\footnote{In the
  $\theta=0$ case, the relationship \eqref{eq:LargeN_neutralrep}
  implies the agreement of not only the profile but also
  the overall normalization of the singular-value spectrum
  for U($N$) and SU($N$)
  at large $N$.
  Note, however, that the relationship \eqref{eq:LargeN_neutralrep}
  does not hold for nonzero $\theta$,
  which allows the appearance of the $\theta$-dependent
  overall factor for U($N$) but not for SU($N$).}.

As an interesting consequence of this fact,
we obtain the free energy
corresponding to \eqref{eq:EK_volume_independent} as
\begin{equation}
  \frac{1}{V} \log Z = C^{(0)} N^2
  - \chi_N \varphi(\theta)
  +  \frac{1}{V}\log \left( \sum_r  e^{VC_r^{(1)}} \right)
  + \cdots \ ,
  \label{eq:EK_volume_independent-theta}
\end{equation}
where the ellipsis represents the terms that vanish in the large-$N$ limit.
Thus we find
\begin{equation}
\frac{1}{V} {\rm Im}\langle  Q \rangle
=  - \frac{1}{V} \frac{d}{d\theta}  \log Z(\theta)
=   \chi_N \frac{d\varphi(\theta)}{d\theta} 
+ \cdots \ ,
  \label{eq:EK_volume_independent2}
\end{equation}
which is independent of $V$ at large $N$.
This explains our observation in Fig.~\ref{fig:ImQLargeN}.
Note that the volume independence we find here
goes beyond the Eguchi-Kawai reduction
in that it refers to the O(1) terms in
\eqref{eq:EK_volume_independent-theta}
instead of the O($N^2$) term.
It should
  be noted here that the leading O($N^2$) term
  in the free energy cannot depend on $\theta$ since
  the $\theta$ term is sub-dominant to the kinetic term of the gauge field
  in the action at large $N$.

Let us also emphasize that
\eqref{eq:EK_volume_independent-theta}
does not hold
in the weak coupling phase $\lambda<2$ since the representations
with all the charge $q$ survive in that case,
and they have a factor $e^{-\frac{1}{2} \chi_N (\theta+2\pi q)^2}$,
which depends on $q$.
Hence, this volume independence holds only for $\lambda>2$.





\section{Summary}
\label{SummaryandOutlook}

In this paper, we discussed how the TRG can be applied to
SU($N$) and U($N$) gauge theories
by considering the 2D case, which is exactly solvable.
Using the character expansion, the fundamental tensor turns out to be
extremely simple as in \eqref{eq:fundamental_tensor},
which results in a simple evolution of the singular-value spectrum with coarse graining.
However, in actual calculation,
we have to restrict the number of representations in the
character expansion
in order to obtain the singular-value spectrum correctly up to a given $D_\text{cut}$.
We have shown that this can be done efficiently by truncating the representations
by their dimensionality using the fact that the totally symmetric representation gives the
smallest dimensionality among the representations with the same $l_1$.

We have investigated the properties
of the singular-value spectrum thus obtained
in detail.
In particular, we found that the spectrum
has a definite profile in the large-$N$ limit up to an overall factor.
This led us to
a new interpretation of the Eguchi-Kawai reduction
in the large-$N$ limit based on
the self-similarity of the tensor network under the course-graining procedure.
We also found an interesting change in the large-$N$ behavior of the spectrum
in the $\text{U}(N)$ case at the
critical
coupling
of the Gross-Witten-Wadia phase transition.
As a result, the singular-value spectrum obtained
in the strong coupling phase
becomes identical to that in the $\text{SU}(N)$ case in the large-$N$ limit.

The singular values obtained up to large $D_\text{cut}$
enabled us to obtain explicit results for finite $N$ and finite $V$,
and we discussed, in particular, 
how the Gross-Witten-Wadia third-order phase transition appears as $N$ increases.
We also investigated
the U($N$) gauge theory with
the $\theta$ term,
and showed how the first-order phase transition at $\theta=\pi$ associated with
the spontaneous breaking of parity symmetry appears as the volume increases.
In the large-$N$ limit, we observed a new type of volume independence
for the topological charge in the strong coupling phase.
We provided a theoretical understanding of this property
by investigating the large-$N$ behavior of the singular-value spectrum
in the presence of the $\theta$ term.
This volume independence
goes beyond the Eguchi-Kawai reduction in that it refers to the O(1) term
of the free energy.
Whether this phenomenon occurs in large-$N$ gauge theories other
than the ones considered in this paper is an interesting open question.

On the technical side, our work clearly shows that the TRG is indeed potentially
useful in investigating
SU($N$) and U($N$) gauge theories in general.
However, we should also be aware of various
technical complications that appear in applying the method to
theories
in higher dimensions and/or with matter fields.
Note, in particular, that the singular values are determined not only by the representations
but also by the internal degrees of freedom introduced by
the
Clebsch-Gordan coefficients \cite{Bazavov2019}.
The practical strategy to restrict the number of
representations proposed in this paper
is expected to be useful even in that case.


\subsection*{Acknowledgements}
We would like to thank
D.~Kadoh for valuable discussions.
The computations were carried out on PC clusters in KEK Computing Research Center
and KEK Theory Center.

\appendix

\section{Proof for the representation with the minimal dimensionality}
\label{sec:proof}

In this section, we prove that the representation of the SU($N$) group
with the minimal dimensionality for a given $l_1$ is
the totally symmetric representation
and its complex conjugate.
Given that this fact plays a crucial role in our proposal for restricting the number
of representations in Section \ref{chooserep},
we provide the proof in an elementary fashion so that it is accessible
to all the interested readers.

Let us recall that the
representation of the SU($N$) group is specified by
\eqref{UNrepresentations}
with an extra constraint $l_N =0$,
and the dimensionality $d_r$ is given by \eqref{eq:representation_dimensionality}.
The statement is that $d_r$
becomes minimum for a given $l_1 = \Lambda$
when $l_2 = \cdots = l_{N-1}  = 0$ or $\Lambda$,
which we prove by induction.

Let us first take the log of \eqref{eq:representation_dimensionality} as
\begin{equation}
  \log   d_r=\sum_{1\leq i<j\leq N}
  \log \left(1+\frac{ l_i- l_j}{j-i}\right) \ .
  \label{eq:representation_dimensionality_log}
\end{equation}
We can easily check that the statement is true for
the $N=3$ case, where we only have to minimize 
\eqref{eq:representation_dimensionality_log}
with respect to $l_2$ for fixed $l_1=\Lambda$ and $l_3=0$.
We obtain
\begin{equation}
  \log   d_r= \log (\Lambda+1-l_2) (1+ l_2)  \ ,
  \label{eq:representation_dimensionality_log-su3}
\end{equation}
which becomes minimum at $l_2 = 0$ and $\Lambda$.

Next we assume that the statement is true for $N$
and prove that it is true also for $N+1$.
The crucial point to note here is that the dimensionality $d_{r}(N+1)$
for the representation $r=\{ l_j\}$ ($j=1,\cdots , N+1$) of the SU($N+1$) group
is given in terms of 
the dimensionality $d_{\tilde{r}}(N)$
for the representation $\tilde{r}=\{ l_{j+1} \}$ ($j=1,\cdots , N$) of the SU($N$) group as
\begin{equation}
  \log   d_r(N+1)= \log d_{\tilde{r}}(N)
 +  \sum_{j=2}^{N}
 \log \left(1+\frac{ \Lambda- l_j}{j-1}\right) + \log \left( 1 + \frac{\Lambda}{N} \right)   \ .
  \label{eq:representation_dimensionality_log-N}
\end{equation}
We minimize \eqref{eq:representation_dimensionality_log-N}
in two steps.
First we fix $l_2=\tilde{\Lambda}$ and minimize 
\eqref{eq:representation_dimensionality_log-N} within this constraint;
then we minimize \eqref{eq:representation_dimensionality_log-N}
with respect to $\tilde{\Lambda}$.

In the first step, we consider the first term and the second term
separately. The first term is minimized when $l_j=0$ or $\tilde{\Lambda}$ for $2 \le j \le N$
by assumption. The second term is minimized when $l_j=\tilde{\Lambda}$ ($2 \le j \le N$). 
Thus we find that \eqref{eq:representation_dimensionality_log-N} is minimized
when $l_j=\tilde{\Lambda}$ ($2 \le j \le N$) for fixed $l_2=\tilde{\Lambda}$.
Plugging this in \eqref{eq:representation_dimensionality_log-N}, we obtain
\begin{equation}
  \log   d_r(N+1)= 
 \sum_{j=2}^{N}
 \log \left(1+\frac{ \tilde{\Lambda}}{j-1}\right) \left(1+\frac{ \Lambda - \tilde{\Lambda}}{j-1} \right) \ .
  \label{eq:representation_dimensionality_log-N-2}
\end{equation}

The second step is to minimize \eqref{eq:representation_dimensionality_log-N-2}
with respect to $\tilde{\Lambda}$.
Similarly to \eqref{eq:representation_dimensionality_log-su3}
for the SU(3) case,
we find that each term in the sum appearing in \eqref{eq:representation_dimensionality_log-N-2}
is minimized when $\tilde{\Lambda}=0$ or $\Lambda$.
Thus we find that the statement is true for $N+1$.
This concludes the proof of the statement for arbitrary $N$ by induction.


\bibliographystyle{JHEP}
\bibliography{TRG}
\end{document}